\pdfoutput=1
\documentclass[11pt,a4paper]{article}
\usepackage{jheppub}
\usepackage{microtype}
\usepackage{dcolumn}
\usepackage{booktabs}

\usepackage{lineno}  

\usepackage{xspace}

\newcommand{\Ba}[1]{\ensuremath{^{#1}\mathrm{Ba}}\xspace}
\newcommand{\Bi}[1]{\ensuremath{^{#1}\mathrm{Bi}}\xspace}
\newcommand{\Co}[1]{\ensuremath{^{#1}\mathrm{Co}}\xspace}

\newcommand{\K}[1]{\ensuremath{^{#1}\mathrm{K}}\xspace}
\newcommand{\Kr}[1]{\ensuremath{^{#1}\mathrm{Kr}}\xspace}

\newcommand{\Pb}[1]{\ensuremath{^{#1}\mathrm{Pb}}\xspace}
\newcommand{\Po}[1]{\ensuremath{^{#1}\mathrm{Po}}\xspace}

\newcommand{\Rn}[1]{\ensuremath{^{#1}\mathrm{Rn}}\xspace}
\newcommand{\Th}[1]{\ensuremath{^{#1}\mathrm{Th}}\xspace}
\newcommand{\Tl}[1]{\ensuremath{^{#1}\mathrm{Tl}}\xspace}
\newcommand{\U}[1]{\ensuremath{^{#1}\mathrm{U}}\xspace}
\newcommand{\Xe}[1]{\ensuremath{^{#1}\mathrm{Xe}}\xspace}

\newcommand{\bbnonu}{\ensuremath{0\nu\beta\beta}\xspace}
\newcommand{\bbtwonu}{\ensuremath{2\nu\beta\beta}\xspace}
\newcommand{\bb}{\ensuremath{\beta\beta}\xspace}
\newcommand{\Qbb}{\ensuremath{Q_{\beta\beta}}\xspace}

\begin{document}

\title{Measurement of radon-induced backgrounds in the NEXT double beta decay experiment}

\collaboration{The NEXT Collaboration}
\author[18]{P.~Novella,}
\author[18]{B.~Palmeiro,}
\author[18,6]{A.~Sim\'on,}
\author[18,a]{M.~Sorel,\note[a]{Corresponding author.}}
\author[10]{C.~Adams,}
\author[14,8]{P.~Ferrario,}
\author[18,19]{G.~Mart\'inez-Lema,}
\author[3,14]{F.~Monrabal,}
\author[16]{G.~Zuzel,}
\author[14,8,b]{J.J.~G\'omez-Cadenas,\note[b]{NEXT Co-spokesperson.}}
\author[18]{V.~\'Alvarez,}
\author[6]{L.~Arazi,}
\author[4]{C.D.R~Azevedo,}
\author[2]{K.~Bailey,}
\author[20]{F.~Ballester,}
\author[18]{J.M.~Benlloch-Rodr\'{i}guez,}
\author[12]{F.I.G.M.~Borges,}
\author[18]{A.~Botas,}
\author[18]{S.~C\'arcel,}
\author[18]{J.V.~Carri\'on,}
\author[21]{S.~Cebri\'an,}
\author[12]{C.A.N.~Conde,}
\author[18]{J.~D\'iaz,}
\author[5]{M.~Diesburg,}
\author[12]{J.~Escada,}
\author[20]{R.~Esteve,}
\author[18]{R.~Felkai,}
\author[11]{A.F.M.~Fernandes,}
\author[11]{L.M.P.~Fernandes,}
\author[4]{A.L.~Ferreira,}
\author[11]{E.D.C.~Freitas,}
\author[14]{J.~Generowicz,}
\author[7]{A.~Goldschmidt,}
\author[19]{D.~Gonz\'alez-D\'iaz,}
\author[10]{R.~Guenette,}
\author[9]{R.M.~Guti\'errez,}
\author[2]{K.~Hafidi,}
\author[1]{J.~Hauptman,}
\author[11]{C.A.O.~Henriques,}
\author[9]{A.I.~Hernandez,}
\author[19]{J.A.~Hernando~Morata,}
\author[20]{V.~Herrero,}
\author[2]{S.~Johnston,}
\author[3]{B.J.P.~Jones,}
\author[18]{M.~Kekic,}
\author[17]{L.~Labarga,}
\author[18]{A.~Laing,}
\author[5]{P.~Lebrun,}
\author[18]{N.~L\'opez-March,}
\author[9]{M.~Losada,}
\author[11]{R.D.P.~Mano,}
\author[10]{J.~Mart\'in-Albo,}
\author[18]{A.~Mart\'inez,}
\author[3]{A.~McDonald,}
\author[11]{C.M.B.~Monteiro,}
\author[20]{F.J.~Mora,}
\author[18]{J.~Mu\~noz Vidal,}
\author[18]{M.~Musti,}
\author[18]{M.~Nebot-Guinot,}
\author[3,c]{D.R.~Nygren,\note[c]{NEXT Co-spokesperson.}}
\author[5]{A.~Para,}
\author[18,d]{J.~P\'erez,\note[d]{Now at Laboratorio Subterr\'aneo de Canfranc, Spain.}}
\author[3]{F.~Psihas,}
\author[18]{M.~Querol,}
\author[18]{J.~Renner,}
\author[2]{J.~Repond,}
\author[2]{S.~Riordan,}
\author[15]{L.~Ripoll,}
\author[18]{J.~Rodr\'iguez,}
\author[3]{L.~Rogers,}
\author[18]{C.~Romo-Luque,}
\author[12]{F.P.~Santos,}
\author[11]{J.M.F. dos~Santos,}
\author[13,e]{C.~Sofka,\note[e]{Now at University of Texas at Austin, USA.}}
\author[13]{T.~Stiegler,}
\author[20]{J.F.~Toledo,}
\author[14]{J.~Torrent,}
\author[4]{J.F.C.A.~Veloso,}
\author[13]{R.~Webb,}
\author[13,f]{J.T.~White,\note[f]{Deceased.}}
\author[18]{N.~Yahlali}
\emailAdd{sorel@ific.uv.es}
\affiliation[1]{
Department of Physics and Astronomy, Iowa State University, 12 Physics Hall, Ames, IA 50011-3160, USA}
\affiliation[2]{
Argonne National Laboratory, Argonne, IL 60439, USA}
\affiliation[3]{
Department of Physics, University of Texas at Arlington, Arlington, TX 76019, USA}
\affiliation[4]{
Institute of Nanostructures, Nanomodelling and Nanofabrication (i3N), Universidade de Aveiro, Campus de Santiago, Aveiro, 3810-193, Portugal}
\affiliation[5]{
Fermi National Accelerator Laboratory, Batavia, IL 60510, USA}
\affiliation[6]{
Nuclear Engineering Unit, Faculty of Engineering Sciences, Ben-Gurion University of the Negev, P.O.B. 653, Beer-Sheva, 8410501, Israel}
\affiliation[7]{
Lawrence Berkeley National Laboratory (LBNL), 1 Cyclotron Road, Berkeley, CA 94720, USA}
\affiliation[8]{
Ikerbasque, Basque Foundation for Science, Bilbao, E-48013, Spain}
\affiliation[9]{
Centro de Investigaci\'on en Ciencias B\'asicas y Aplicadas, Universidad Antonio Nari\~no, Sede Circunvalar, Carretera 3 Este No.\ 47 A-15, Bogot\'a, Colombia}
\affiliation[10]{
Department of Physics, Harvard University, Cambridge, MA 02138, USA}
\affiliation[11]{
LIBPhys, Physics Department, University of Coimbra, Rua Larga, Coimbra, 3004-516, Portugal}
\affiliation[12]{
LIP, Department of Physics, University of Coimbra, Coimbra, 3004-516, Portugal}
\affiliation[13]{
Department of Physics and Astronomy, Texas A\&M University, College Station, TX 77843-4242, USA}
\affiliation[14]{
Donostia International Physics Center (DIPC), Paseo Manuel Lardizabal, 4, Donostia-San Sebastian, E-20018, Spain}
\affiliation[15]{
Escola Polit\`ecnica Superior, Universitat de Girona, Av.~Montilivi, s/n, Girona, E-17071, Spain}
\affiliation[16]{
M. Smoluchowski Institute of Physics, Jagiellonian University, Krakow, 30-348, Poland}
\affiliation[17]{
Departamento de F\'isica Te\'orica, Universidad Aut\'onoma de Madrid, Campus de Cantoblanco, Madrid, E-28049, Spain}
\affiliation[18]{
Instituto de F\'isica Corpuscular (IFIC), CSIC \& Universitat de Val\`encia, Calle Catedr\'atico Jos\'e Beltr\'an, 2, Paterna, E-46980, Spain}
\affiliation[19]{
Instituto Gallego de F\'isica de Altas Energ\'ias, Univ.\ de Santiago de Compostela, Campus sur, R\'ua Xos\'e Mar\'ia Su\'arez N\'u\~nez, s/n, Santiago de Compostela, E-15782, Spain}
\affiliation[20]{
Instituto de Instrumentaci\'on para Imagen Molecular (I3M), Centro Mixto CSIC - Universitat Polit\`ecnica de Val\`encia, Camino de Vera s/n, Valencia, E-46022, Spain}
\affiliation[21]{
Laboratorio de F\'isica Nuclear y Astropart\'iculas, Universidad de Zaragoza, Calle Pedro Cerbuna, 12, Zaragoza, E-50009, Spain}

\abstract{The measurement of the internal \Rn{222} activity in the NEXT-White detector during the so-called Run-II period with \Xe{136}-depleted xenon is discussed in detail, together with its implications for double beta decay searches in NEXT. The activity is measured through the alpha production rate induced in the fiducial volume by \Rn{222} and its alpha-emitting progeny. The specific activity is measured to be $(38.1\pm 2.2~\mathrm{(stat.)}\pm 5.9~\mathrm{(syst.)})$~mBq/m$^3$. Radon-induced electrons have also been characterized from the decay of the \Bi{214} daughter ions plating out on the cathode of the time projection chamber. From our studies, we conclude that radon-induced backgrounds are sufficiently low to enable a successful NEXT-100 physics program, as the projected rate contribution should not exceed 0.1~counts/yr in the neutrinoless double beta decay sample.}
\maketitle

\clearpage

\section{\label{sec:intro}Introduction}

\begin{figure}[!htb]
  \begin{center}
    \includegraphics[width=0.70\textwidth]{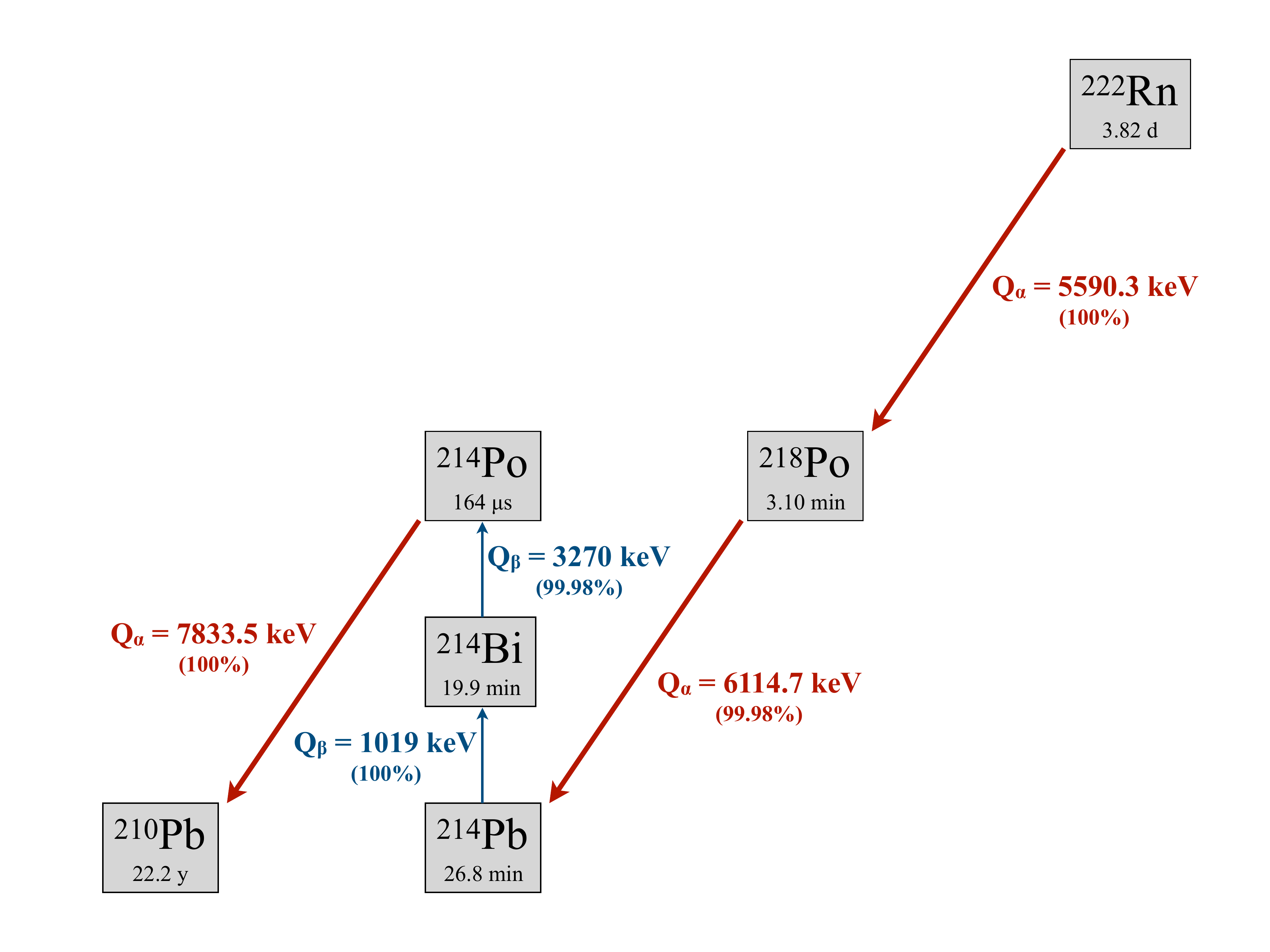} \hfill 
    \caption{\label{fig:radondecayscheme}Part of the \Rn{222} decay scheme that is most relevant for this analysis. Quoted half-lives, Q-values and branching ratios are extracted from the ENSDF database \cite{Wu:2009lpp,Jain:2006yav,ShamsuzzohaBasunia:2014yyr}.}
  \end{center}
\end{figure}

Radon (\Rn{220} and particularly \Rn{222}) has proven to be a serious concern for underground experiments searching for rare events, such as neutrinoless double beta decay (\bbnonu), dark matter interactions, or solar neutrino interactions. The $\alpha$/$\beta$/$\gamma$ radiation and nuclear recoils produced in the decays of \Rn{222} and its progeny are potential backgrounds to these searches. In particular, the interactions of the high-energy (up to 3.18~MeV \cite{Wu:2009lpp}) gamma-rays produced in \Bi{214} $\beta$ decays can be a significant background in \bbnonu experiments. For this reason, all \bbnonu experiments require monitoring and mitigation of radon content. A schematic of the middle part of the naturally-occurring uranium (\U{238}) decay chain, starting from \Rn{222} and ending in the long-lived \Pb{210} isotope, is shown in Fig.~\ref{fig:radondecayscheme}. Similarly, \Rn{220} is part of the thorium (\Th{232}) decay chain, which also includes several $\beta$ decays of relevance for rare event searches. Air-borne radon is present in the atmosphere surrounding the detector. Internal radon can also be present within the detector volume, either via emanation from detector materials, or through air leaks from the surroundings. 

Despite involving the same radioactive isotopes, radon-induced backgrounds often require a special treatment compared to backgrounds from \U{238}/\Th{232} radioactive impurities trapped in detector components. The reason has to do with their different spatial distributions, their unique sensitivity to electrostatic fields, and with the different background mitigation strategies that are at hand. Radon is a highly diffusive and soluble noble gas. Therefore, both air-borne radon external to the detector and internal radon tend to have homogeneous spatial distributions within gaseous/liquid media, unlike radioactive impurities concentrated in solid materials. While radon gas is neutral and uniformly distributed, the daughter isotopes produced in the decay are often positively-charged ions. Hence, the presence of electrostatic fields in the experiments can alter greatly the spatial distribution of radon-induced backgrounds, and plate-out of radon daughters on detector surfaces is common. Finally, radon-induced backgrounds differ from other backgrounds because they can be reduced via active filtration systems. Radon can be removed from gases (including air) using purification columns filled with adsorbers, such as charcoal, typically operated at cryogenic temperatures \cite{Wojcik:2017vux,Benato:2017kdf}. Radon can also be removed via nitrogen gas stripping \cite{Wojcik:2017vux,Benziger:2007aq} or via liquid distillation \cite{Rupp:2017zcy}. Because of these reasons, radio-assay measurements of detector components via gamma-ray spectroscopy or mass spectrometry techniques prior to detector installation are of limited use to constrain radon-induced backgrounds. Highly-sensitive radon emanation tests and/or {\it in-situ} background measurements during detector operations are necessary to understand radon activity.

A high-pressure xenon gas time projection chamber (TPC) is the detector technology chosen by the Neutrino Experiment with a Xenon TPC (NEXT) to search for \bbnonu in \Xe{136} \cite{Alvarez:2012flf}. The detector operates with xenon gas enriched in the \Xe{136} isotope at 10--15~bar pressure. This detector concept features excellent energy resolution. At the Q-value of the \bb reaction, \Qbb=(2457.83 $\pm$ 0.37)~keV \cite{Redshaw:2007un}, the resolution is about 1\% FWHM \cite{Lorca:2014sra,Renner:2018ttw}. This is possible thanks to the electroluminescent-based readout of the experiment, that is a nearly noiseless amplification stage for the ionization signal. In addition, the identification of the double electron topological signature characteristic of \bbnonu is possible thanks to the detector low-density and fine spatial granularity of the tracking readout \cite{Ferrario:2015kta}. Finally, this technology shows promise to effectively detect the \Ba{136} daughter produced in a \Xe{136} \bb decay \cite{McDonald:2017izm}. This additional handle may provide background-free conditions for \bb detectors at the ton-scale and beyond. After a successful prototyping period in 2008--2014 \cite{Alvarez:2012yxw,Alvarez:2013gxa,Gonzalez-Diaz:2015oba}, the experiment has moved to underground and radio-pure operations with the NEXT-White detector at the Laboratorio Subterr\'aneo de Canfranc (LSC). The next phase of the experiment is called NEXT-100, and will make use of 100~kg of \Xe{136}-enriched xenon, also at the LSC. The sensitivity of NEXT-100 to \Xe{136} \bbnonu decay has been evaluated in \cite{Martin-Albo:2015rhw}, relying on detailed radio-assay measurements \cite{Alvarez:2012as,Alvarez:2014kvs,Cebrian:2017jzb} and Monte-Carlo simulations. A background index of $4\times 10^{-4}$~counts/(keV$\cdot$kg$\cdot$~yr) at most is expected in the \bbnonu energy region of interest after all cuts, yielding a sensitivity of $6\times 10^{25}$~yr after an exposure of 275~kg$\cdot$yr. For the 29~keV wide energy region of interest and the 91~kg \Xe{136} active mass considereed in \cite{Martin-Albo:2015rhw}, this corresponds to a background rate of $\lesssim 1.06$~counts/yr. This sensitivity study assumed a negligible contribution from radon-induced backgrounds, at the $\lesssim 0.03$~counts/yr rate level. The purpose of this work is to test this assumption. In particular, we focus here on internal radon within the xenon recirculation loop of NEXT, as air-borne radon external to the detector is expected to be very effectively mitigated in the experiment.

A previous study of radon-induced backgrounds that is particularly relevant to this work, for the similarities with NEXT in the experimental goals and techniques employed, is that of the EXO-200 \bbnonu experiment. The EXO-200 Collaboration reports a background rate of 0.24~counts/yr in the 150~keV wide \bbnonu energy region of interest (ROI) from \Rn{222} decays inside the liquid xenon \cite{Albert:2015nta}. This rate is attributed to radon emanation from either external xenon piping or from internal readout cables. The EXO-200 Collaboration also installed a charcoal-based deradonator to suppress, down to negligible levels, air-borne radon from the air gap between the experiment's copper cryostat and lead shield \cite{Albert:2017owj}. This rate of \bbnonu background induced by internal radon was found to be largely negligible in EXO-200, as it corresponds to about 1\% of the total background budget \cite{Albert:2017owj}. However, a similar rate could be significant in NEXT-100, aiming for much lower background conditions. Indeed, the $\lesssim 1.06$~counts/yr total background rate estimated for NEXT-100 is of the same order of the EXO-based estimate of radon-induced backgrounds alone, hence the importance of a direct measurement of radon-induced backgrounds in NEXT.

This paper is organized as follows. Section~\ref{sec:nextwhite} gives a description of the NEXT-White detector and its operating conditions. Section~\ref{sec:radonalphas} studies the radon-induced alpha particles produced in NEXT-White. This alpha production rate is used as a basis for the radon activity measurement presented in Sec.~\ref{sec:radonactivity}. Section~\ref{sec:radonelectrons} focuses on radon-induced beta/gamma activity in NEXT-White. Implications for double beta decay searches in NEXT are discussed in Sec.~\ref{sec:betabetabackgrounds}.

\section{\label{sec:nextwhite}The NEXT-White detector at the LSC}

The NEXT detection concept \cite{Alvarez:2012flf} is based on a high-pressure gaseous xenon TPC. NEXT uses electroluminescence (EL) as a nearly noiseless amplification stage for ionization produced in the xenon gas. The EL (also called secondary, or S2) scintillation light is used for separated energy and tracking measurements. The light is read by two planes of photo-detectors located at opposite ends of the detector cylindrical structure. The energy plane is located behind the transparent cathode, and detects the backward EL light using photomultiplier tubes (PMTs). The tracking plane is located a few mm away from the EL gap, and detects the forward EL light using silicon photomultipliers (SiPMs). The energy plane sensors detect also the primary (or S1) scintillation light produced promptly in the active volume, for event $t_o$ determination. The detector triggers on energy plane information, either S1 or S2 light.

The NEXT-White (NEW) detector\footnote{Named after Prof. James White, our late mentor and friend.} is the currently operating realization of this detector concept. A detailed description of NEXT-White can be found in \cite{Monrabal:2018xlr}. The detector is located at the Laboratorio Subterr\'aneo de Canfranc (LSC, Spain). The detector active volume is 530.3~mm long along the drift direction, and has a 198~mm radius. The EL gap is 6~mm wide. The energy plane read-out consists of 12 Hamamatsu R11410-10 PMTs, providing 31\% coverage. Also relevant to this analysis is the 130~mm long xenon gas buffer volume separating the TPC cathode from the energy plane. The tracking plane read-out consists of 1792 SensL C-Series SiPMs, placed on a 2D lattice at 10~mm pitch. The detector was commissioned in late 2016. Extensive calibration data were taken during NEXT-White Run-II, in 2017. Some low-background data were also taken during Run-II, and are the subject of this work. The main scientific goals of NEXT-White, to be undertaken in 2018--2019, are the full validation of the NEXT background model and the measurement of the \Xe{136} \bbtwonu decay mode.

The Run-II of NEXT-White lasted from March 21st, 2017 to November 20th, 2017. All data were taken with xenon depleted in the \Xe{136} isotope. The gas pressure was set to approximately 7.2~bar for most of Run-II data, corresponding to a xenon mass in the active volume of about 2.6~kg. The standard conditions for drift and EL field were $\simeq$0.4~kV/cm and $\simeq$1.7~kV/(cm$\cdot$bar), respectively \cite{Monrabal:2018xlr}. The electron drift velocity was accurately measured to be 0.97~mm/$\mu$s for these standard Run-II conditions, see \cite{Simon:2018vep}. For the dedicated radon-induced alpha runs discussed in Secs.~\ref{sec:radonalphas} and \ref{sec:radonactivity}, lower EL field configurations of order $\simeq$0.7~kV/(cm$\cdot$bar) were chosen to avoid saturation of the photo-detector signals. The gas purity continuously improved during Run-II, from $\simeq$150~$\mu$s electron lifetime at the beginning, to $\simeq$1,800~$\mu$s towards the end \cite{Monrabal:2018xlr}. All Run-II data were taken with a 6~cm thick copper shield within the pressure vessel. An additional, 20~cm thick, shield structure made of lead bricks surrounding the detector was used during the second part of Run-II. All data discussed in this work were taken without the lead shield. A variety of radioactive sources were used during Run-II for calibration purposes \cite{Monrabal:2018xlr}. While none of the radon-induced data discussed in this work used decays from these sources, some of the runs were taken concurrently with sources near/inside the detector. In this case, trigger and offline event selection were used to effectively reject calibration source events.

Particularly relevant to this paper is the choice of getters to purify the gas. All data presented in this work were taken with the heated getter-based purifier MonoTorr PS4-MT50-R-535 from SAES. However, during Run-II, the observed radon emanation rates were affected by periods of operation with gas circulating through an ambient temperature purifier, in the few days or weeks preceding the heated getter runs. The ambient temperature getter used during Run-II is the MicroTorr MC4500-902FV model from SAES. As is well known, see for example \cite{Calvo:2016hve} and Sec.~\ref{sec:radonalphas} in this work, the radon emanation of this ambient temperature getter cartridge is intolerably high for rare event searches. Long-term operations with heated getters only during the second half of 2017 and 2018 have demonstrated that the use of room temperature getters are in fact not necessary to purify the xenon gas in NEXT-White to sufficient levels. Low radon conditions as discussed in Sec.~\ref{sec:radonalphas} (period A1) can be reached concurrently with long ($>2$~ms) electron lifetime conditions in NEXT-White.

The measured activity of airborne radon (\Rn{222}) at the Laboratorio Subterr\'aneo de Canfranc (Hall A) varies between 60 and 80~Bq~m$^{-3}$ \cite{Bandac:2017eks}. Left at this level, airborne radon would represent an intolerably high source of gamma-rays from \Bi{214}, at the level of $10^{-3}$~counts/(keV$\cdot$kg$\cdot$~yr), see \cite{Martin-Albo:2015rhw}. For this reason, a radon abatement system by ATEKO A.S. has been installed in Hall A at the LSC. Radon-free air will be flushed into the air volume enclosed by the lead castle starting in 2018, thanks to a controllable air delivery system built for this purpose. Measurements of the \Rn{222} content in the air delivered by the pipes and reaching the lead castle give $<1.5$~mBq/m$^3$, that is, a radon reduction of 4-5 orders of magnitude compared to Hall A air. While the radon activity in the air surrounding the pressure vessel still needs to be measured, we expect air-borne radon contributions to the background budget to be completely negligible in NEXT.

\section{\label{sec:radonalphas}Radon-induced alpha particles}

The detector operating conditions for the radon-induced alpha runs are summarized in Tab.~\ref{tab:alpharunconditions}. Three alpha run periods are considered. The first period (A1) corresponds to a time when the ambient temperature getter had not yet been turned on. This is the period used to estimate the \Rn{222} internal activity for the upcoming physics runs in NEXT-White and NEXT-100. The second period (A2) was taken shortly after the ambient temperature getter was operated for the first time, during April 11 -- 21, 2017. A number of short runs were taken over two weeks, to measure electron lifetime and to monitor how the high radon activity induced by the getter would decrease over time. The third period (A3) occurred shortly after a second period of operations with the ambient temperature getter (May 18 -- July 12, 2017), again generating a high alpha production rate. The A3 sample is used to relate the \Rn{222}-induced alpha production rate within the xenon fiducial volume with the rate of electrons from \Bi{214} daughter plate-out on the cathode.

\begin{table}[!htb]
\caption{\label{tab:alpharunconditions}Detector operating conditions during the NEXT-White Run-II alpha runs. The dates refer to the 2017 calendar year. The voltage drops across the drift and EL regions are given by $\Delta V_\mathrm{drift}$ and $\Delta V_\mathrm{EL}$, respectively. The column $\tau_e$ indicates the range of electron lifetimes measured.}
\begin{center}
\begin{tabular}{cccccc}
\hline
Period & Date             & Pressure & $\Delta V_\mathrm{drift}$ & $\Delta V_\mathrm{EL}$ & $\tau_e$  \\
       &                  & (bar)    & (kV)              & (kV)            & ($\mu$s) \\ \hline
A1     & Mar 30 -- Apr 3  & 7.05 -- 7.08    & 18.9          & 3.1             & 220 -- 294 \\ 
A2     & Apr 22 -- May 7  & 6.85 -- 6.88    & 19.2 -- 20.0  & 2.8             & 409 -- 595 \\
A3     & Jul 17 -- Jul 18 & 7.21            & 21.0          & 2.8             & 1061 -- 1084 \\ \hline
\end{tabular}
\end{center}
\end{table}

The voltage drop across the electroluminescent (EL) region, $\Delta V_\mathrm{EL}$, was set in the range 2.8--3.1~kV. Considering a 6~mm nominal EL region thickness and the gas pressures indicated in Tab.~\ref{tab:alpharunconditions}, the reduced EL field during the A1--A3 runs is $E_\mathrm{EL}/P$=0.65--0.73~kV/(cm$\cdot$bar). For comparison, the secondary scintillation threshold in pure xenon gas at 10~bar reported in \cite{Oliveira:2011xk}, from a fit to simulation results, is $(E_\mathrm{EL}/P)_\mathrm{thr}$=0.69--0.76~kV/cm. Hence, the EL settings empirically chosen to avoid signal saturation appear to be very close to the EL threshold reported in the literature. The EL gain, that is the number of S2 photons produced per ionization electron reaching the EL gap, is very sensitive to $E_\mathrm{EL}/P$ conditions when operating near threshold. On the other hand, the drift field conditions, in the $E_\mathrm{drift}$=0.36--0.39~kV/cm range, provided a stable electron drift velocity. The electron drift velocity is accurately measured run-by-run, and found to be 0.95--0.97~mm/$\mu$s, by using alpha particles emitted from the cathode. All alpha runs discussed in this work, except one special run taken to enable trigger efficiency studies, relied on the same S2 trigger configuration. The alpha S2 trigger was set to search for high-charge and point-like energy depositions, as is the case for alpha particles produced in the gas. Approximately ten thousand triggers were taken during A1, while the A2 and A3 periods accounted for a few hundred thousand triggers overall. Table~\ref{tab:alpharunconditions} also reports the electron lifetime measured during these runs, obtained by fitting the exponential attenuation of charge as a function of drift distance. The electron lifetime ranges span both time variations during those periods, as well as the percent-level fit uncertainties. The electron lifetime continuously improved from $\simeq$0.25~ms during A1 to $\simeq$1.1~ms during A3. 

Three event reconstruction steps are run on raw data. First, binary data are converted into PMT and SiPM waveforms in HDF5 format. Second, waveforms are processed to zero-suppress the data and find S1/S2 peaks. Third, a point-like event reconstruction is performed. A 5.5~MeV kinetic energy alpha particle has a CSDA range of 3.5~mm at 7~bar. Hence, the point-like event reconstruction is a very good approximation in the case of alpha particles. After this step, alpha events are characterized by a single (X,Y,Z) position, where Z is the drift direction, as well as S1 and S2 yields in photoelectrons (PEs). The drift distance Z is obtained from the drift time $t_\mathrm{S2}-t_\mathrm{S1}$ as measured by the PMTs, times the electron drift velocity measured with cathode alphas. The (X,Y) positions are obtained from a barycenter of the SiPM charges.

\begin{figure}[!htb]
  \begin{center}
    \includegraphics[width=0.49\textwidth]{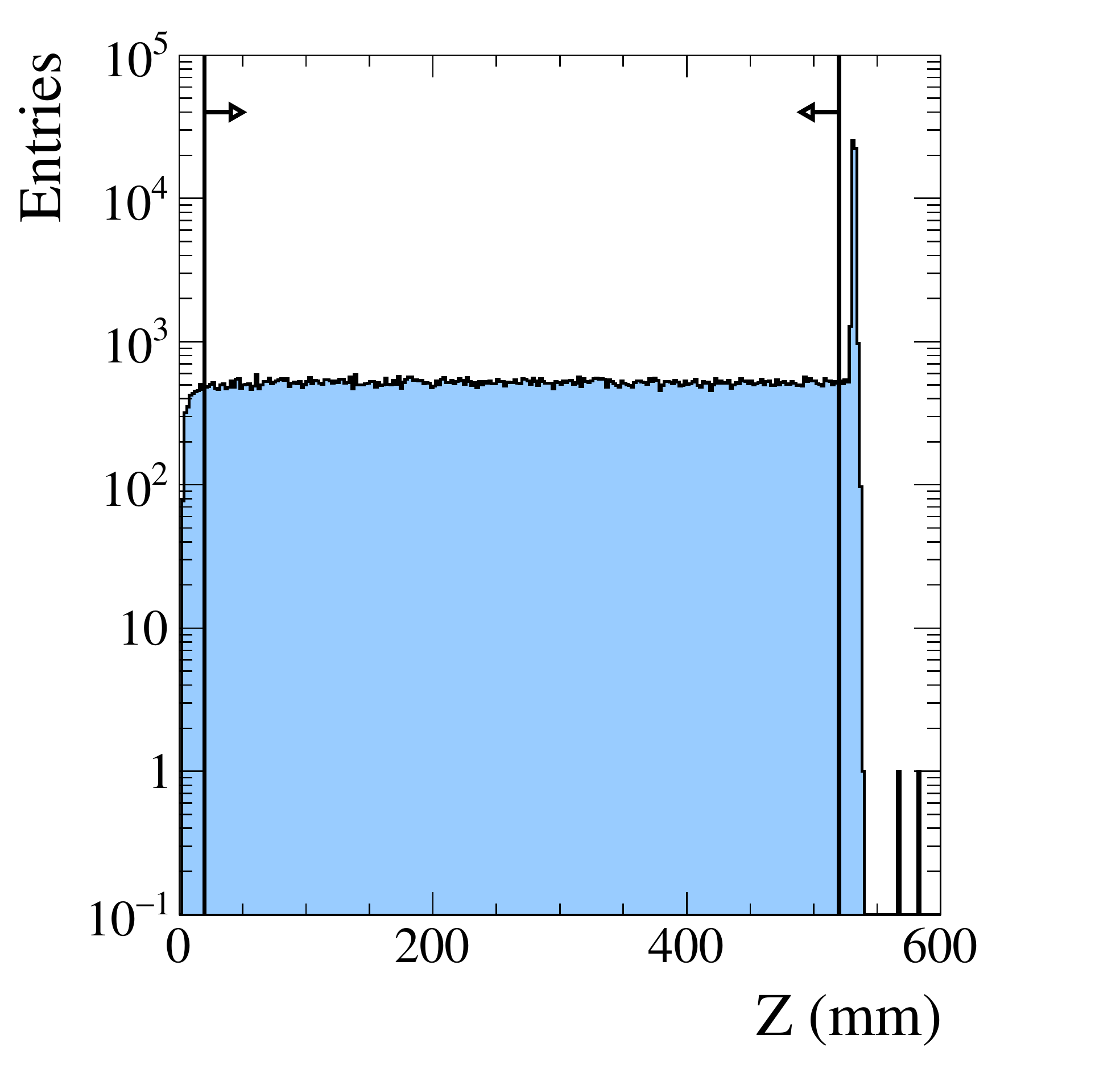} \hfill
    \includegraphics[width=0.49\textwidth]{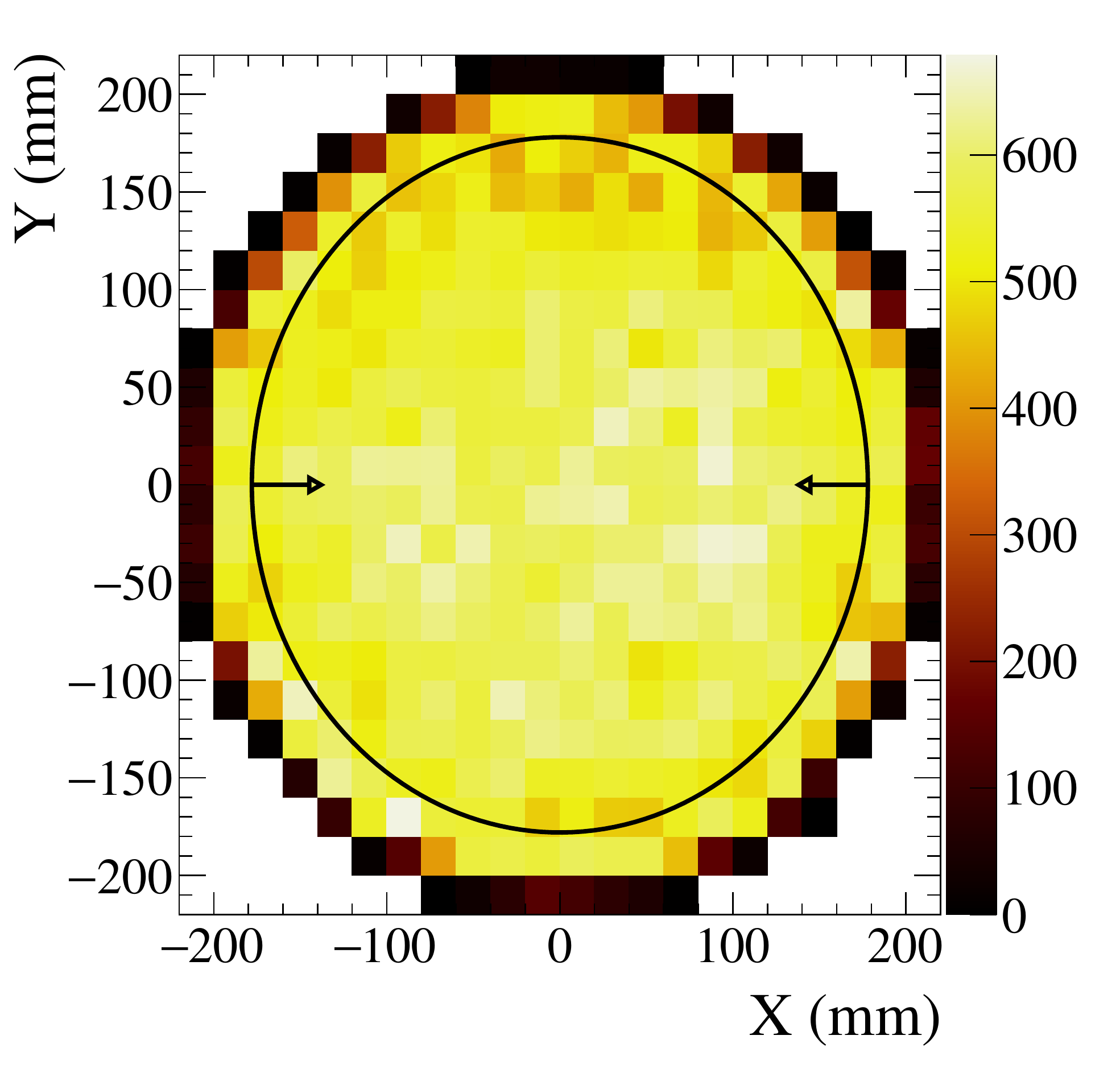} \hfill
    \caption{\label{fig:clusterpositionfilter}Spatial distribution of alpha candidate events. Left panel: Z distribution. Right panel: distribution in the (X,Y) plane. In the two panels, the solid black lines and arrows indicate the fiducial region.}
  \end{center}
\end{figure}

In order to isolate an {\it inclusive alpha} candidate sample, a S1/S2 peak selection on the PMT waveforms is performed. We require one and only one alpha-like S1 peak, and one and only one alpha-like S2 peak, per event. An alpha-like S1 or S2 peak is easily identified in the data based on peak charge, peak time width and peak start time information, similarly to what was done previously for NEXT-DEMO data \cite{Alvarez:2012hu}. The spatial distribution of inclusive alpha candidates during the A3 period is shown in Fig.~\ref{fig:clusterpositionfilter}. The distribution is homogeneous within the entire active volume $0<Z<530.3$~mm, $R\equiv \sqrt{X^2+Y^2}<198$~mm, with the exception of a clear excess of alpha particles produced from the cathode plane at $Z=530.3$~mm. Cathode alphas are produced from the plate-out and subsequent decay of alpha-emitting radon daughters on the cathode. Successfully reconstructed cathode alphas represent a 23--28\% fraction of the inclusive alpha sample. 

\begin{figure}[!htb]
  \begin{center}
    \includegraphics[width=\textwidth]{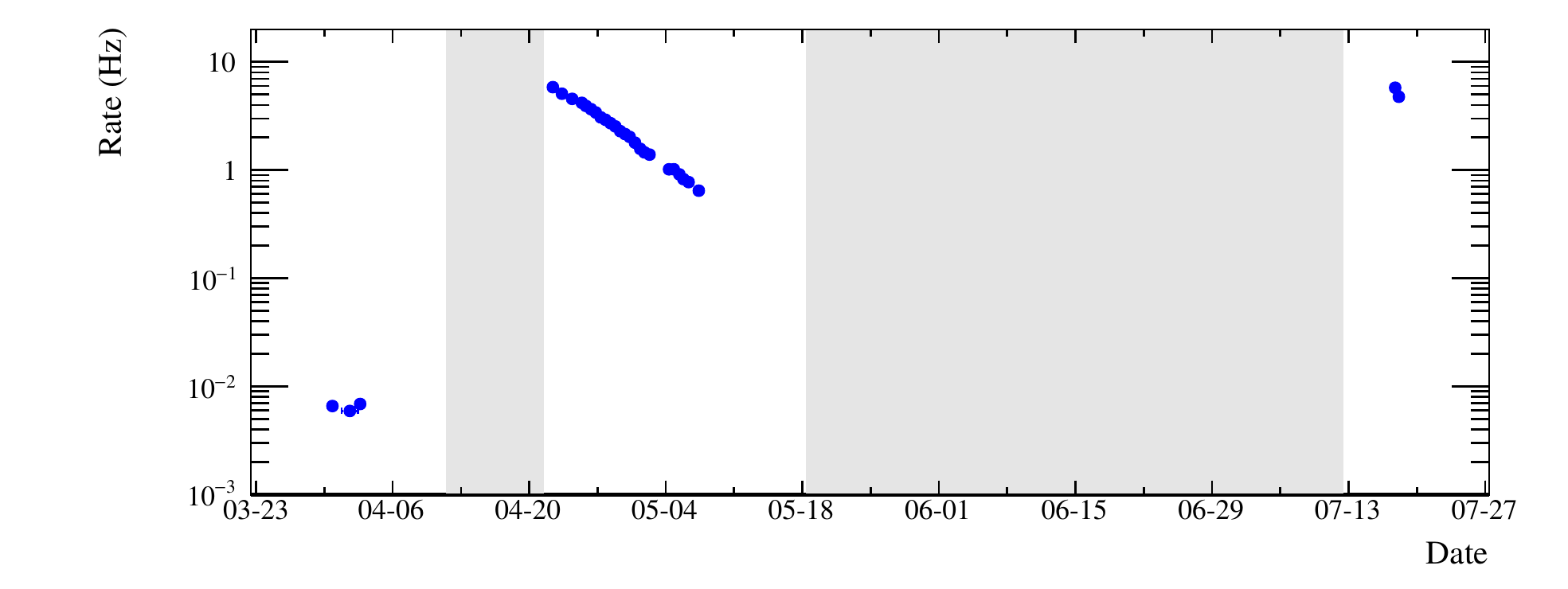}
    \caption{\label{fig:alphasrate}Alpha production rate during run periods A1--A3. Grey areas indicate periods when the ambient temperature getter was operating.}
  \end{center}
\end{figure}

The time evolution of the inclusive alpha production rate is shown in Fig.~\ref{fig:alphasrate}. The grey areas indicate the time when the ambient temperature getter was operating, hence introducing high concentrations of radon into the gas system and the detector. The inclusive alpha production rate is found to be constant at about 6~mHz during the three runs forming A1. The rate is much higher, of the order of 1~Hz or more, during run periods A2 and A3. The inclusive alpha production rate decreases over time during A2, as time passes since ambient temperature getter operations.

\begin{figure}[!htb]
  \begin{center}
    \includegraphics[width=0.7\textwidth]{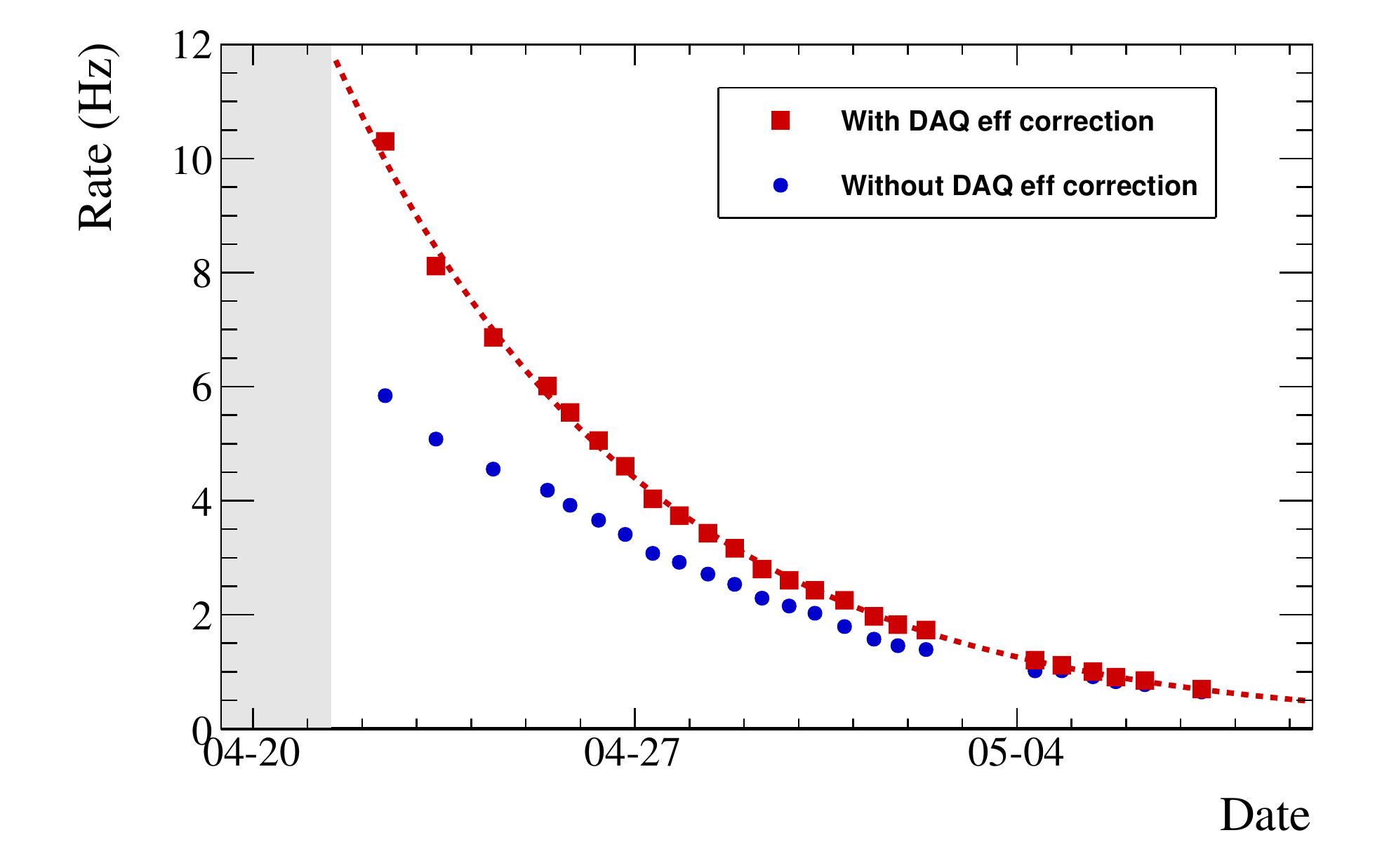}
    \caption{\label{fig:radondecay}Radon decay evolution during period A2, with and without DAQ efficiency correction. An exponential fit is superimposed.}
  \end{center}
\end{figure}

Figure~\ref{fig:radondecay} shows with greater detail the time evolution of the inclusive alpha production rate during A2. DAQ dead-time effects are relevant, considering that the trigger rate was in the several Hz range and a trigger mask was set to 15~Hz during these runs. A 15~Hz trigger mask means that the DAQ cannot acquire additional events during a 1/15~s = 66.7~ms time period following a trigger. Given the known average trigger for each run, one can therefore estimate the DAQ efficiency as follows:
\begin{equation}
\varepsilon_\mathrm{DAQ}=1-R_\mathrm{trg}/R_\mathrm{max}
\label{eq:effdaq}
\end{equation}
\noindent where $R_\mathrm{trg}$ is the average trigger rate measured, and $R_\mathrm{max}$ is the trigger mask. The inclusive alpha production rate both with and without the DAQ efficiency correction of Eq.~\ref{eq:effdaq} is shown in Fig.~\ref{fig:radondecay}. The efficiency-corrected alpha production rate is fitted as a function of time via an exponential plus a fixed ($6.2\times 10^{-3}$~Hz) constant term. The latter has been estimated from the rate measured during the A1 period. The DAQ efficiency correction greatly improves the quality of the fit, also shown in the figure. The exponential half-life returned by the fit over the entire A2 period is $T_{1/2} = (3.871\pm 0.013)$~d. This value is close to the \Rn{222} half-life reported in the literature, $T_{1/2}(\Rn{222})=(3.8235\pm 0.0003)$~d \cite{Jain:2006yav}, confirming that alpha particles are induced by \Rn{222} decays.

\section{\label{sec:radonactivity}Radon activity measurement}

The time evolution of the alpha production rate during the A2 period unambiguously identifies those particles as being induced from the decay of \Rn{222} and its progeny. However, this information is not sufficient to disentangle the relative contributions of the \Rn{222} (5590~keV), \Po{218} (6115~keV) and \Po{214} (7834~keV) alpha emitters in the chain, see Tab.~\ref{fig:radondecayscheme}, and hence to provide an absolute radon activity measurement\footnote{In the alpha decays listed above and throughout this work, the visible energy values quoted in parenthesis match the Q-value of the reaction. This is about 100~keV higher than the alpha kinetic energy, as it includes also the contribution from the nuclear recoil kinetic energy. The latter is also visible in a fully active detector such as NEXT.}. In addition, other alpha emitters might be present, for example from the decay of \Rn{220} and its daughters. Because of these reasons, a spectroscopic analysis has been performed, with the goal of identifying the alpha-emitting isotopes. In order to reconstruct the energy more reliably, a {\it fiducial alpha} candidate sub-sample within the inclusive sample has been defined. This selection is illustrated in Fig.~\ref{fig:clusterpositionfilter} by the solid black lines. The fiducial volume is defined to be $20 < Z < 520$~mm and $R<178$~mm, corresponding to a 2.0~kg xenon fiducial mass at the 7.2~bar operating gas pressure. About 54\% of the alpha candidates pass the fiducial requirement.

Our procedure to define the alpha energy estimator is similar to the one employed for NEXT-DEMO data, see \cite{Serra:2014zda}. The procedure accounts for $(X,Y,Z)$ spatial inhomogeneities in the detector response, and combines S1 and S2 information to reduce the impact of electron-ion recombination fluctuations on energy resolution. The alpha energy calibration procedure is done in four steps. First, the $Z$ dependence of S1 and S2 yields is corrected. In a second step, the $(X,Y)$ dependence of the $Z$-corrected S1 and S2 yields is accounted for. In order to mitigate recombination fluctuations, we define the alpha energy estimator as follows:
\begin{equation}
E\equiv \lambda (N_1+N_2/\eta)
\label{eq:es1s2}
\end{equation}
\noindent where $N_1$ and $N_2$ are the $(X,Y,Z)$-corrected S1 and S2 yields, respectively, $\eta$ is a weight factor to rescale the ionization yield component, and $\lambda$ is an overall conversion factor. In a third step we find the optimum $\eta$ value, that is the one providing the best relative energy resolution of the \Rn{222} peak, according to the energy estimator of Eq.~\ref{eq:es1s2}. Once the optimum $\eta$ value has been found, in a fourth and final step we determine the overall conversion factor from PEs/PMT to keV energy units. This is the $\lambda$ factor in Eq.~\ref{eq:es1s2}. This factor is obtained by aligning the fitted \Rn{222} peak position with 5590.3~keV.

\begin{figure}[!htb]
  \begin{center}
    \includegraphics[width=\textwidth]{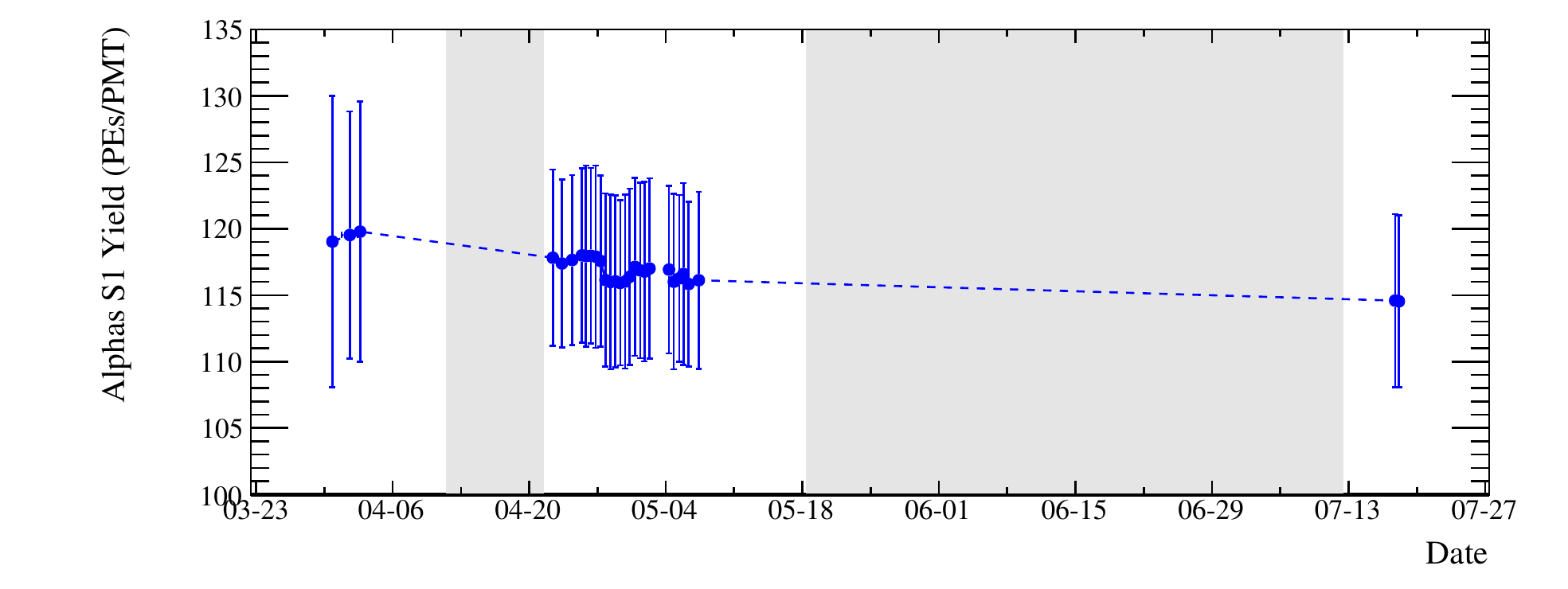}
    \includegraphics[width=\textwidth]{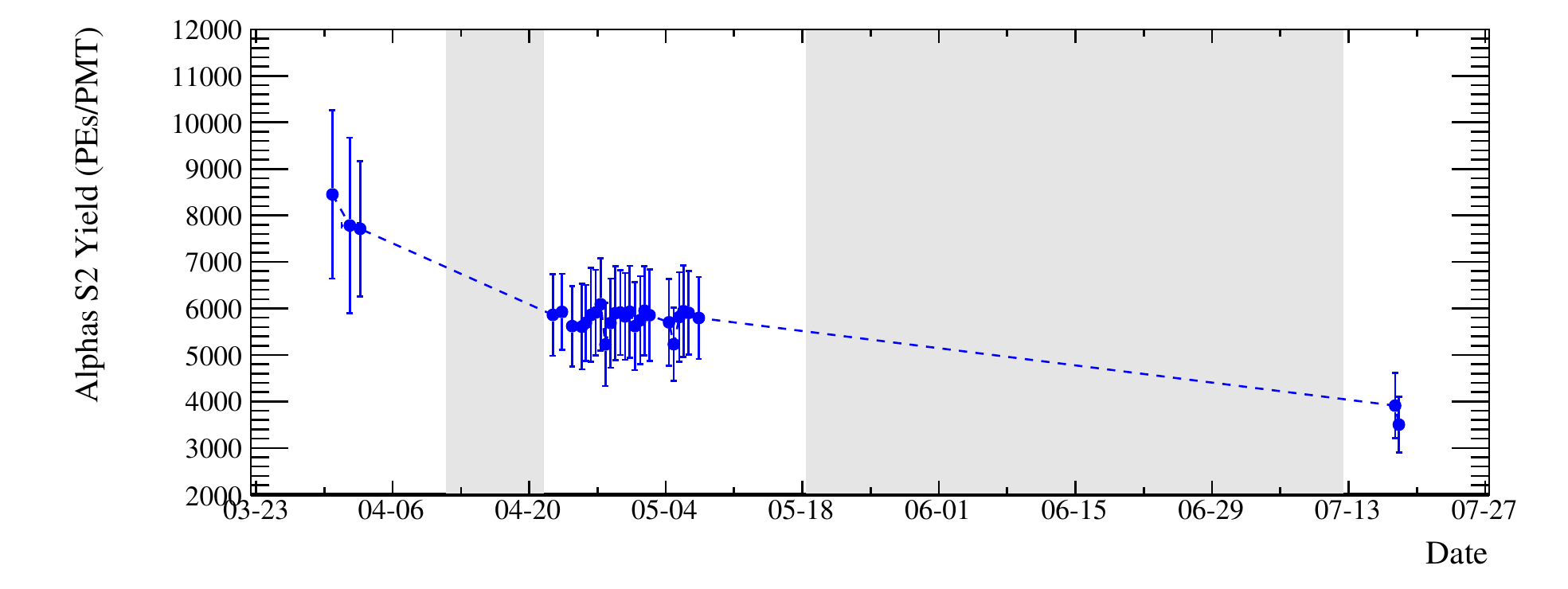}
    \caption{\label{fig:alphass1s2yields}S1 (top) and S2 (bottom) yields, corrected for the $Z$ spatial dependence and for all fiducial alpha events, during run periods A1--A3. The data points and error bars indicate the mean and the RMS of the $Z$-corrected distributions, respectively.}
  \end{center}
\end{figure}

Figure~\ref{fig:alphass1s2yields} shows the average S1 and S2 yields for fiducial alpha particles as a function of time. The time evolution is shown after $Z$ corrections, mainly to account for the time-varying electron lifetime. The average alpha S1 yield is stable to within 4\% during the entire A1--A3 runs, hence providing a robust (although not very accurate) energy estimate. On the other hand, the average S2 yield decreases by more than a factor of two over the same period. This is not surprising, since even small changes in the reduced EL field $E_\mathrm{EL}/P$ (as the ones inferred from Tab.~\ref{tab:alpharunconditions}) can give rise to large variations in the S2 yield, when operating the detector near EL threshold conditions. For this reason, the energy scale $\lambda$ in our energy estimator of Eq.~\ref{eq:es1s2} is separately computed for each run period.

\begin{figure}[!htb]
  \begin{center}
    \includegraphics[width=0.49\textwidth]{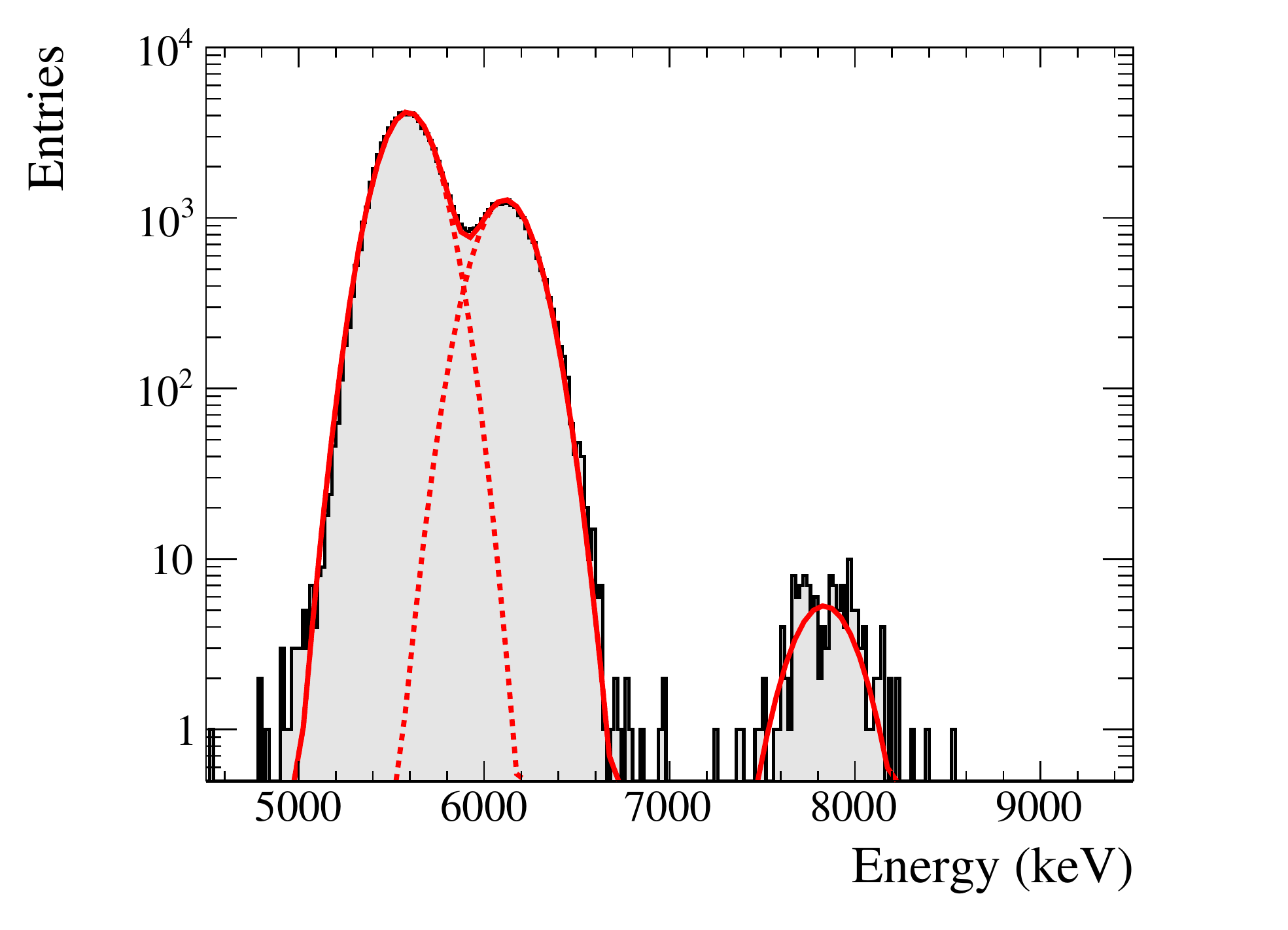} \hfill
    \includegraphics[width=0.49\textwidth]{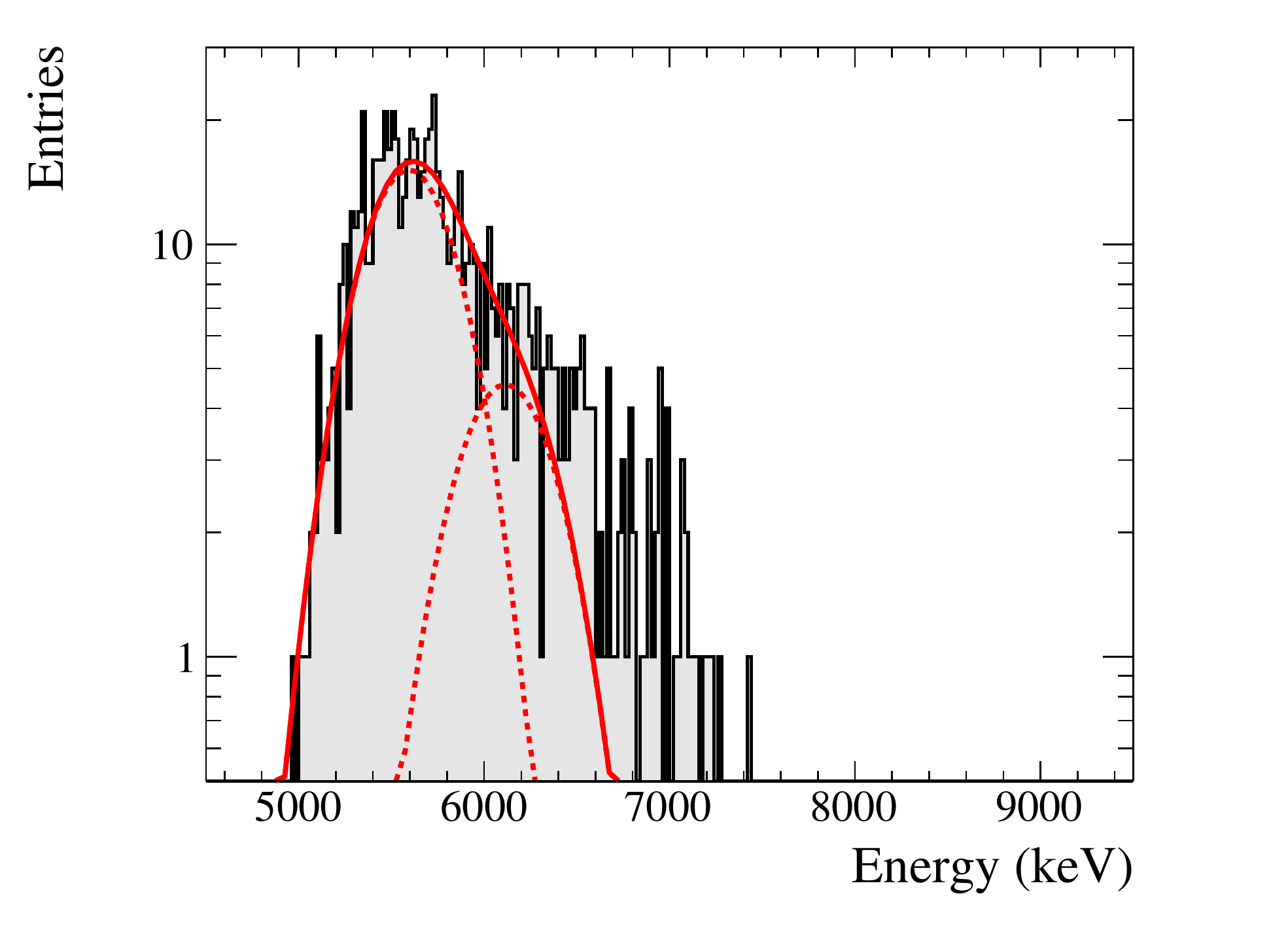} \hfill 
    \caption{\label{fig:alphaenergy}Energy distribution for fiducial alpha candidate events during run periods A3 (left panel) and A1 (right panel). A triple gaussian fit is superimposed to the A3 histogram to describe the \Rn{222} (5590~keV), \Po{218} (6115~keV) and \Po{214} (7834~keV) populations. Only the \Rn{222} and \Po{218} yields are fitted in the A1 histogram.}
  \end{center}
\end{figure}

The alpha energy spectrum obtained during the high (A3) and low (A1) radon activity periods is shown in Fig.~\ref{fig:alphaenergy}. For both periods, the \Rn{222} and \Po{218} populations are described by gaussian distributions. For A3, an excess of events compatible with \Po{214} (7834~keV) is also visible and fitted with a third gaussian. For the lower statistics A1 period, no \Po{214} contribution is visible. Only the overall normalizations of the alpha-emitting isotopes, plus the peak position and width of the \Rn{222} population, are kept free in the fit. The peak position and width of the \Po{218} and \Po{214} populations are rescaled from the corresponding fit parameters of the \Rn{222} population, taking into account the known alpha decay Q-values.

\begin{table}[!htb]
\caption{\label{tab:radonactivity}Fitted yields and specific \Rn{222} activity in the fiducial volume of NEXT-White, during run periods A1 and A3.} 
\begin{center}
\begin{tabular}{ccccc}
\hline
Period & \Rn{222} events   & \Po{218} events & \Po{214} events & \Rn{222} activity \\
       &                   &                 &                 & (Bq/m$^3$)  \\ \hline
A1     & $487\pm 28$   & $154\pm 21$   & - & $(38.1\pm 2.2){\times}10^{-3}$ \\ 
A3     & $(7.28\pm 0.03){\times}10^4$ & $(2.32\pm 0.02){\times}10^4$ & $(1.09\pm 0.11){\times}10^2$ & $37.56\pm 0.14$ \\ \hline
\end{tabular}
\end{center}
\end{table}

The results of the fits shown in Fig.~\ref{fig:alphaenergy} are reported in Tab.~\ref{tab:radonactivity}. For the high radon period A3, the fitted \Rn{222}, \Po{218} and \Po{214} yields represent a $(75.4\pm 0.3)$\%, $(24.1\pm 0.2)$\% and $(0.113\pm 0.011)$\% fraction of the observed fiducial alpha events, respectively. The sum of the fitted yields are therefore compatible with 100\%. For the low radon period A1, the fitted \Rn{222} and \Po{218} yields represent a $(62.7\pm 3.6)$\% and $(19.8\pm 2.7)$\% fraction of the observed fiducial alpha events, respectively. In the latter case, about a 17\% fraction of observed alpha events is unaccounted for by the fit model. The specific \Rn{222} activity measured in the detector fiducial volume is also reported in Tab.~\ref{tab:radonactivity}, by dividing the fitted \Rn{222} yields by the fiducial volume ($4.98\times 10^{-2}$~m$^3$) and by the run durations. A specific \Rn{222} activity of $(38.1\pm 2.2)$ mBq/m$^3$ within the detector fiducial volume is obtained during A1. The \Rn{222} specific activity increases by three orders of magnitude during A3, taken only six days after closing the \Rn{222}-emanating ambient temperature getter.

From Fig.~\ref{fig:alphaenergy}, we can also conclude that no evidence, not even in trace amounts, for \Rn{220} and its progeny is present during the high \Rn{222} activity A3 period. The main alpha emitters in the \Rn{220} chain are expected to be \Rn{220} (6405~keV) itself and its daughter \Po{216} (6906~keV). We estimate the \Rn{220} activity to be at most a $\sim 10^{-3}$ fraction compared to the \Rn{222} activity during A3. For what concerns the low \Rn{222} activity period A1, the fit model in Fig.~\ref{fig:alphaenergy} does not describe the data well. One possibility is that this is due to alpha energy reconstruction biases and poor energy resolution effects, with no significant alpha emitter present beyond \Rn{222} and \Po{218}. Another possibility is that the data excess over the fit in the 6000--7000~keV region is due to sub-dominant contributions from other alpha emitters such as \Rn{220} and \Po{216}, accounting for the remaining $\simeq$17\% fraction of fiducial alpha candidates. The latter option is consistent with the few percent higher alpha S1 yields observed during A1 compared to A2--A3 periods, see the top panel of Fig.~\ref{fig:alphass1s2yields}. While no definitive statement can be made, it is clear that the dominant contribution to the alpha production rate is due to the decay of \Rn{222} and its progeny also during the low radon period A1.

Possible sources of inefficiency and background contamination in the \Rn{222} fiducial yield measurement during A1 have been quantified. The overall systematic uncertainty in the \Rn{222} activity measurement is estimated to be 16\%, dominated by possible alpha energy mis-reconstruction effects and by event reconstruction inefficiencies. Our result for the \Rn{222} specific activity during the low radon period A1 is therefore $(38.1\pm 2.2\mathrm{(stat.)}\pm 5.9\mathrm{(syst.)})$~mBq/m$^3=(38.1\pm 6.3)$~mBq/m$^3$.

It is also interesting to examine the ratio of \Po{218} to \Rn{222} yields within the fiducial volume. During A3, the ratio is measured to be $(31.92\pm 0.25)$\%, see Tab.~\ref{tab:radonactivity}. The \Po{218} to \Rn{222} ratio during the low \Rn{222} period A1 is compatible with this value, albeit with larger errors. Considering that the alpha decay branching ratios of \Rn{222} and \Po{218} are both essentially 100\% (see Fig.~\ref{fig:radondecayscheme}), the complementary fraction provides a measurement of the \Po{218} ion fraction produced in \Rn{222} alpha decays in gaseous xenon: $(68.08\pm 0.25)$\%. The ejected alpha particle in \Rn{222} decay can free several electrons from the \Po{218} atom, transforming it into a positively-charged ion. \Po{218} ions then plate out on negatively-charged surfaces and outside the detector fiducial volume, particularly on cathode wires, where they decay. On the other hand, neutral \Po{218} atoms can also be obtained if the positively-charged ion is neutralized through electron-ion recombination as it drifts through xenon. By selecting fiducial alpha candidates in the $6,100 < E < 6,600$~keV energy range (see Fig.~\ref{fig:alphaenergy}), we have verified that a fiducial alpha sub-sample enriched in \Po{218} decays is also uniformly distributed in space, as observed for the inclusive alpha sample in Fig.~\ref{fig:clusterpositionfilter}. This is consistent with the hypothesis of a nuclear decay of a neutral \Po{218} atom. It is difficult to predict the proportions of neutral and charged \Po{218} atoms, although one expects the majority of them to be in ionized form in a gaseous detector \cite{Argyriades:2009vq}. Our measured ion fraction confirms this expectation. Our measurement can also be compared with the measurement of the same ion fraction provided by the EXO-200 Collaboration in liquid xenon, $(50.3\pm 3.0)$\% \cite{Albert:2015vma}. The EXO-200 measured fraction is lower, as expected from the higher electron-ion recombination in liquid.

The ratio of \Po{214} to \Po{218} alpha decay activities within the fiducial volume  has also been quantified during the high radon activity period A3. In this case the ratio has been measured to be only $(0.47\pm 0.05)$\%, almost two orders of magnitude lower than the \Po{218} to \Rn{222} alpha decay activity ratio. Owing to the short \Po{214} half-life, this ratio can be used to extract the fraction $f_{\beta}$ of \Bi{214} daughters produced in ionized form in \Pb{214} $\beta$ decays, see Fig.~\ref{fig:radondecayscheme}. Given that the fraction of positively-charged \Pb{214} and \Bi{214} ions that decay inside the fiducial volume while drifting is negligible in xenon gas, the \Po{214} to \Po{218} decay ratio in the fiducial volume can be approximated as \cite{Albert:2015vma}:
\begin{equation}
\frac{N(\Po{214})}{N(\Po{218})} \simeq (1-f_{\beta})\cdot (1-f_{\alpha})
\label{eq:fbeta}
\end{equation}
\noindent where $(1-f_{\alpha})$ is the fraction of neutral \Pb{214} daughters produced in \Po{218} alpha decays. Considering the similarity between the two alpha decays, we take this number to be the same as the fraction of neutral \Po{218} daughters produced in \Rn{222} alpha decays, $(31.92\pm 0.25)$\%. Solving for the \Bi{214} daughter ion fraction in \Pb{214} $\beta$ decays, we obtain $f_{\beta}=(98.53\pm 0.15)\%$, where the error is statistical-only. This ion fraction can be compared with the corresponding one obtained in liquid xenon by the EXO-200 Collaboration, $f_{\beta}=(76.4\pm 5.7)\%$ \cite{Albert:2015vma}. Again, a higher ion fraction is obtained in gaseous xenon, presumably owing to the smaller electron-ion recombination rate. The fact that $f_{\beta}>f_{\alpha}$ can also be explained on the same grounds, as the much higher ionization density present in alpha decays leads to higher recombination. We remark that a daughter ion fraction close to 100\% in single $\beta$ decays in xenon gas detectors is suggestive of a similarly high daughter ion fraction in the double $\beta$ decay of \Xe{136}. If confirmed by future dedicated measurements in NEXT, this would be a very positive result toward an effective tagging of the $\Ba{136}^{++}$ daughter, as pursued by NEXT \cite{McDonald:2017izm}. 

In summary, given that $f_{\beta}\simeq 100\%$, we assume in the following that all of the radon-induced \Bi{214} decays that contribute to double beta decay backgrounds originate from the cathode plane. This is different from the EXO-200 case, where only 58\% of the radon-induced \Bi{214} decays that contribute to \bbnonu backgrounds are expected to originate from the TPC cathode, see \cite{Albert:2015nta}.

\section{\label{sec:radonelectrons}Radon-induced electrons}

During Run-II, background runs with no calibration sources deployed and with nominal electric fields have been taken to study radon-induced electrons. Two periods are considered in this analysis. The first one (E1 or {\it high \Rn{222} activity}) corresponds to a specific date when the ambient temperature getter had been turned off for 5 days, the radon activity being still very high. The second period (E2 or {\it low \Rn{222} activity}) corresponds to the data taken 21 and 22 days after the ambient temperature getter was turned off. 

\begin{table}[tbh]
\caption{\label{tab:bgrunconditions}Detector operating conditions for the NEXT-White electron background runs taken during Run-II. The dates refer to the 2017 calendar year. The voltage drops across the drift and EL regions are given by $\Delta V_\mathrm{drift}$ and $\Delta V_\mathrm{EL}$, respectively. The column $\tau_e$ indicates the range of electron lifetimes measured.}
\begin{center}
\begin{tabular}{cccccc}
\hline
Period & Date                 & Pressure       & $\Delta V_\mathrm{drift}$  & $\Delta V_\mathrm{EL}$ & $\tau_e$ \\
       &                      & (bar)          & (kV)               & (kV)           & ($\mu$s) \\ \hline
E1     & Jul 17               & 7.20           & 21.0               & 7.0            & 1089 -- 1130 \\
E2     & Aug 2 -- Aug 3       & 7.190          & 21.0               & 7.0            & 1289 -- 1365 \\ \hline
\end{tabular}
\end{center}
\end{table}

The corresponding dates and detector operating conditions are summarized in Tab.~\ref{tab:bgrunconditions}. As shown in the table, both the pressure and electric fields (or $\Delta V_\mathrm{drift}$ and $\Delta V_\mathrm{EL}$) were kept constant over the two periods. The gas purity continuously improved during E1-E2, approximately ranging from $1.1$ to $1.3$~ms electron lifetimes. The DAQ and trigger configuration was the same for all runs. The trigger configuration relied on a set of loose cuts, requesting a minimum charge of 2$\times10^{5}$ ADC counts per PMT and a time width between 7 and 250 $\mu$s for the S2 signals. The trigger requirement imposed a minimum threshold on deposited energy of about 500~keV. About $10^5$ triggers were taken for each of the two periods.

As for the alpha runs, the event reconstruction for electron tracks is divided in three steps. However, since the background electrons cannot be treated as point-like energy depositions above a certain energy in 7~bar pressure xenon gas (few hundreds keV), the third reconstruction step differs. In this case, the SiPM clusters providing the $X$ and $Y$ coordinates are reconstructed separately for each time (or $Z$) slice of the S2 signals. Electron-like events are kept by requiring only one S1 signal and only one S2 signal per event, hence suppressing \Bi{214}-\Po{214} delayed coincidences characterized by additional alpha-induced activity from \Po{214} decay, and by requiring the S1 yield not to exceed 82~PEs/PMT, see the top panel of Fig.~\ref{fig:alphass1s2yields} 

\begin{figure}[tbh]
  \begin{center}
    \includegraphics[width=0.50\textwidth]{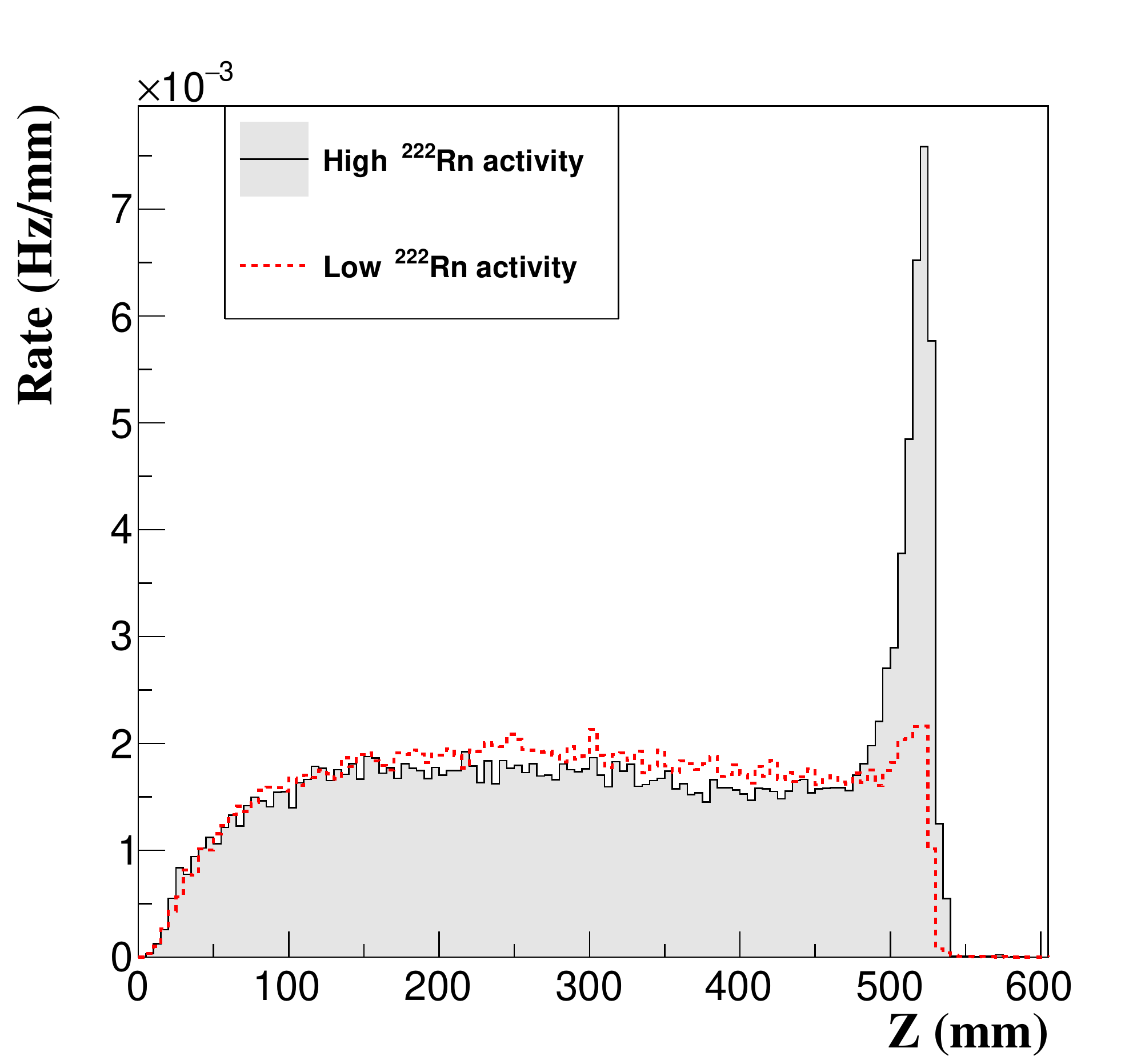} \hfill 
    \caption{\label{fig:ebgrate} Radon-induced electron background rate as a function of $Z$, for the two run periods described in Tab.~\ref{tab:bgrunconditions}.}
  \end{center}
\end{figure}

The electron-like event rate for the two high (E1) and low (E2) radon activity periods and as a function of the $Z$ coordinate is shown in Fig.~\ref{fig:ebgrate}. In the figure, the $Z$ coordinate is defined as the charge-weighted average over all the time slices in the track. The electron-like activity at the cathode position, $Z=530.3$~mm, decreases noticeably as a function of time, as the radon content in the detector decreases. The comparison within the active volume ($Z<530.3$~mm) shows instead a stable rate between the E1 and E2 periods, implying a non-radon origin for those electron tracks. 

\begin{figure}[tbh]
  \begin{center}
    \includegraphics[width=0.99\textwidth]{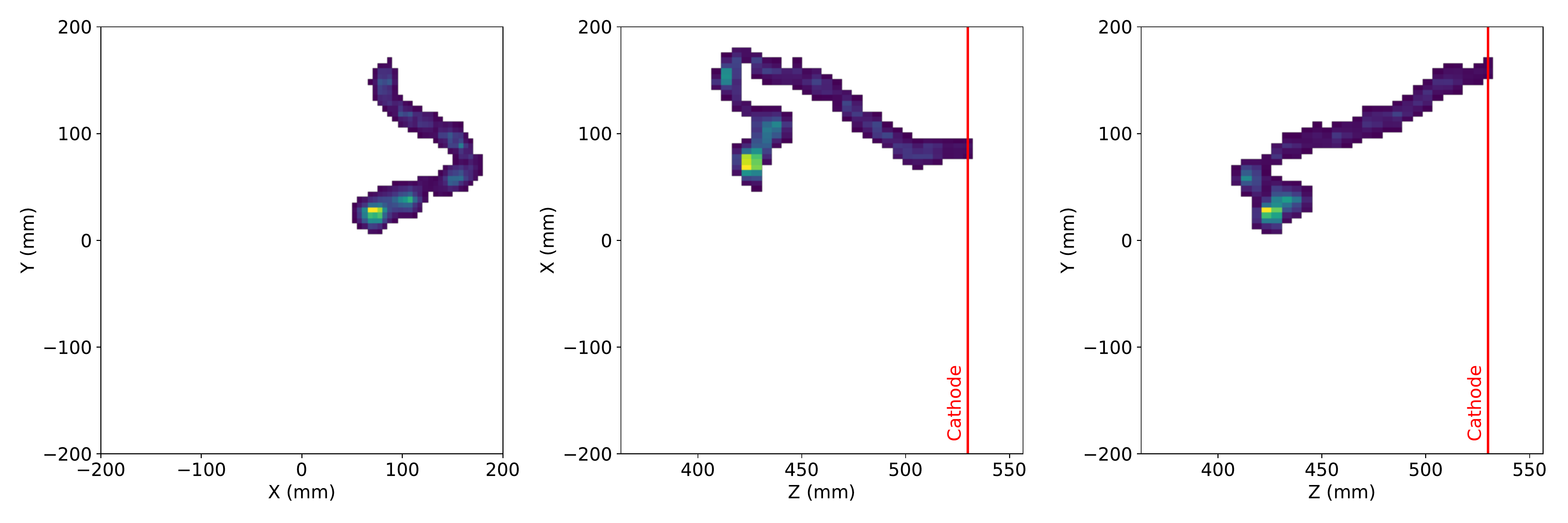} \hfill 
    \caption{\label{fig:ebgtrack} High energy ($E>1.5$~MeV) electron originating from the cathode. The three projections, $XY$ (left), $ZX$ (center) and $ZY$ (right), show the expected topological signature of an electron.}
  \end{center}
\end{figure}

The topology of a typical high-energy ($E>1.5$~MeV) electron track originating from the cathode, from run period E1, is shown in Fig.~\ref{fig:ebgtrack}. The three projections are reconstructed in this case with a Maximum Likelihood Expectation Maximization (ML-EM) algorithm described in \cite{Simon:2017pck}. It can be clearly seen how the track originates within the cathode plane. Also, the track has the typical topological signature of an electron, with an erratic path and a high-energy deposition at the track end-point.

The overall rate and relevant distributions for radon-induced electrons originating from the cathode have been compared between data and Monte-Carlo (MC) expectations. Electron events with $Z_\mathrm{max}>520$~mm, where $Z_\mathrm{max}$ is the the largest Z position among all hits in the event, are taken to be cathode electrons. A pure data sample of radon-induced cathode electrons is obtained by subtracting the E2 (low \Rn{222}) period from the E1 (high \Rn{222}) period. In this statistical subtraction, we take into account the residual radon activity induced by the ambient temperature getter that is still present during E2, by considering the average time difference between the two periods (16.3~days) and the known \Rn{222} half-life (3.82~days). The data rate has been corrected for the DAQ inefficiency, as described by Eq.~\ref{eq:effdaq}.

The corresponding MC sample was obtained by generating $10^7$ \Bi{214} isotropic decays, uniformly distributed in space within the NEXT-White cathode plane. Considering the short half-life of the \Po{214} daughter, 163.6~$\mu$s \cite{ShamsuzzohaBasunia:2014yyr}, the alpha decays of \Po{214} are also simulated with their proper time distribution. The predictions are based on a full simulation. The radioactive decays and the energy deposition within the xenon active volume at 7.2 bar gas pressure are based on a Geant4-based \cite{Agostinelli:2002hh} simulation, see for example \cite{Martin-Albo:2015dza} for details. Electron drift effects including diffusion and attachment, S1/S2 light production and propagation within the detector, and the full electronics response of the energy and tracking readout planes are simulated. The simulation outcome is a set of digitized PMT/SiPM waveforms, as for raw data. The MC sample has been reconstructed following the same procedure applied to the data. The MC normalization has been estimated from the radon specific activity measurement of the A3 alpha sample, $A=(37.56\pm 0.14)$~Bq/m$^3$, see Tab.~\ref{tab:radonactivity}. In order to estimate the total \Bi{214} activity from the cathode, we assume that the relevant \Rn{222} decay volume, the one from which \Bi{214} daughters ultimately plate on the cathode, extends throughout the full active and buffer gas volumes of the detector. In the MC normalization estimate, we also correct for the DAQ efficiency during A3 (see Eq.~\ref{eq:effdaq}) and for the 1.0~day average time difference between the A3 and E1 runs, resulting in more \Rn{222} content during the earlier E1 run compared to A3. As a result, we estimate that $10^7$ simulated \Bi{214} decays correspond to a live-time of ($20.5\pm 5.0$)~days. The error in the MC exposure stems from the uncertainty in the relevant \Rn{222} decay volume to be considered in the calculation.

The radon-induced cathode electron rate measured in data is found to be $(0.096\pm 0.003)$~Hz, to be compared with a MC expectation of $(0.142\pm 0.028)$~Hz. The uncertainty in the data measurement is statisical-only, while a systematic error of 20\% has been assigned to the MC expectation from the uncertainty in the MC exposure mentioned above. The latter error is found to dominate over other systematic error sources, such as data/MC differences in event trigger, reconstruction and selection efficiencies. In other words, our measurement of the \Rn{222}-induced alpha production in the xenon fiducial volume is consistent with the \Bi{214}-induced electron production rate from the cathode within 1.6$\sigma$. Alpha data are therefore a useful tool to understand the latter background source in NEXT.

\begin{figure}[tbh]
  \begin{center}
    \includegraphics[width=0.49\textwidth]{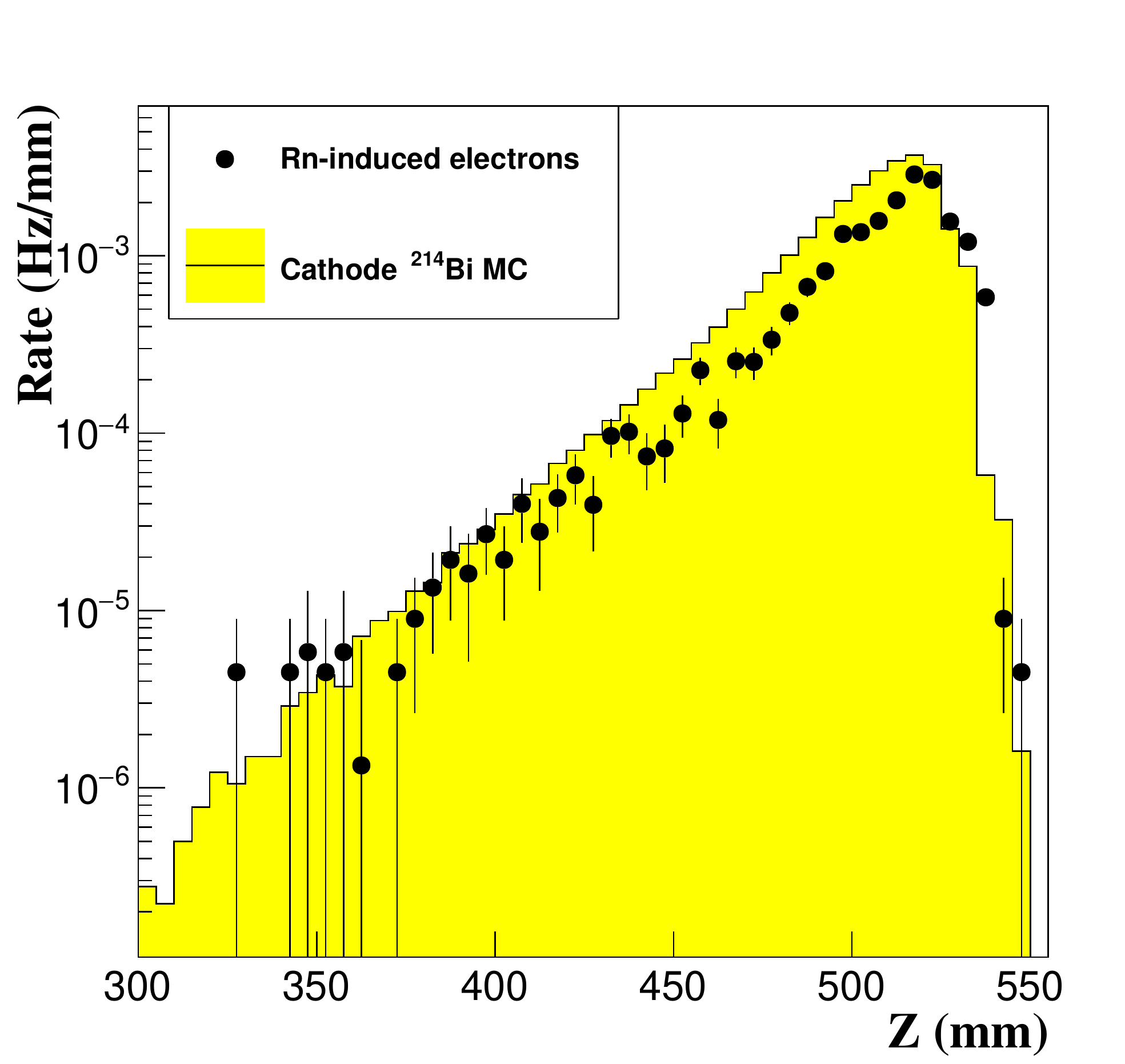} \hfill
    \includegraphics[width=0.49\textwidth]{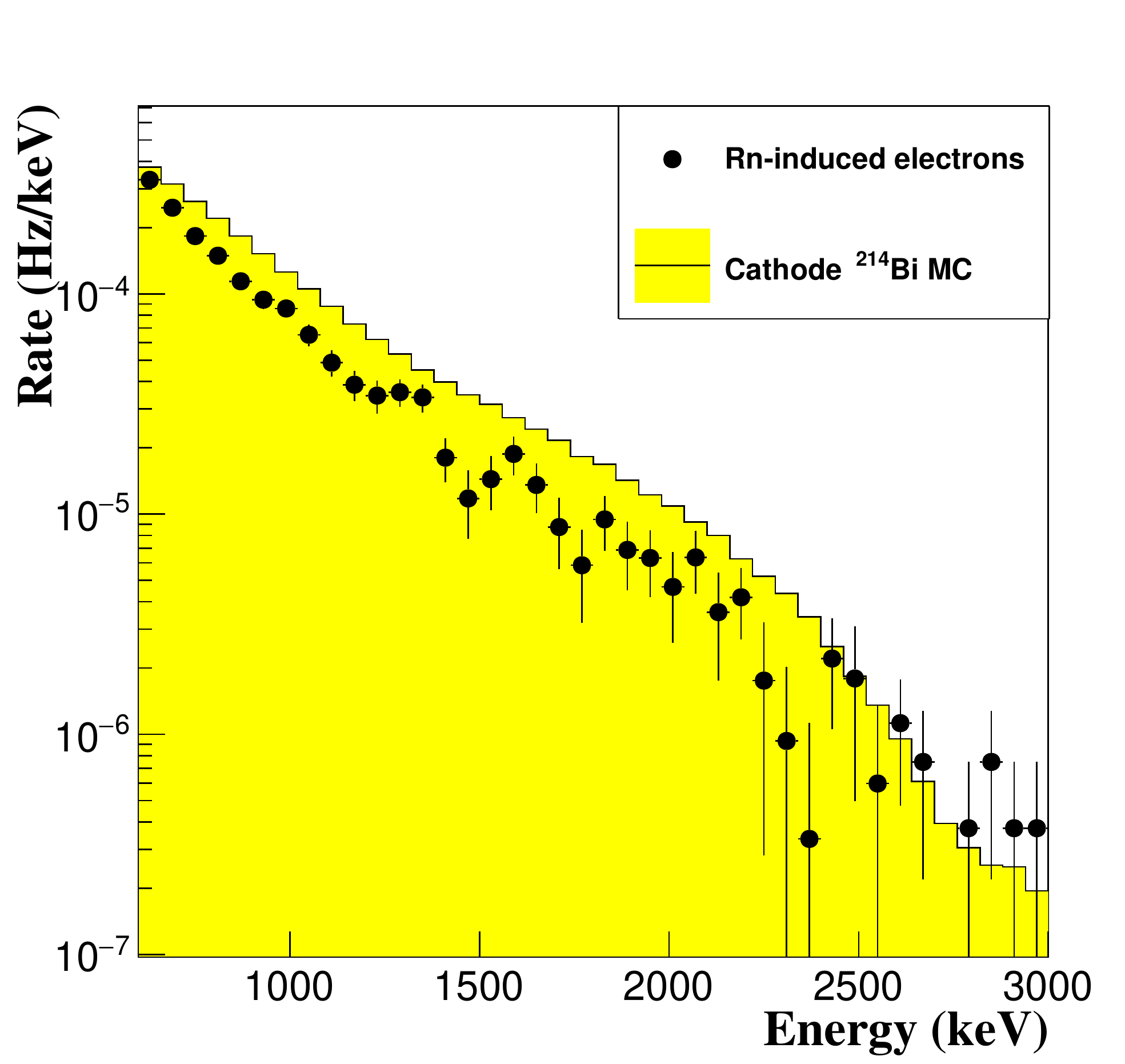}    
    \caption{\label{fig:ebgmc} Radon-induced electrons originating from the cathode (black dots) superimposed on the MC expectation (yellow histogram), as a function of the reconstructed $Z$ coordinate (left panel) and reconstructed energy (right).}
  \end{center}
\end{figure}

Figure~\ref{fig:ebgmc} shows the reconstructed $Z$ and energy distributions of radon-induced cathode electrons in data and MC. Both data and MC rates are absolutely normalized. The $Z$ distribution shows a clear peak at the cathode position, $Z=530.3$~mm, for both data and MC. The asymmetry in the peak toward $Z<530.3$~mm is due to the significant track extent within the active volume, and to the procedure to define an effective $Z$ position per event from the charge-weighting of all the time slices. 

The energy of cathode electron events has also been reconstructed, and is shown in the right panel of Fig.~\ref{fig:ebgmc}. The electron energy reconstruction includes three steps. The steps are similar to the ones described in Sec.~\ref{sec:radonactivity} for the alpha energy estimate, but they are generalized to extended tracks and they rely on the S2 charge only. First, the energy associated to each reconstructed hit in the event is separately corrected for electron attachment. The electron lifetime assumed for the correction was measured with alpha runs. The second step consists of a geometrical $XY$ correction of the detector response depending on the hit XY position. The correction relies on a $XY$ energy map obtained from a \Kr{83m} calibration run taken about two weeks prior to the E1 data, see \cite{Martinez-Lema:2018ibw} for details. Finally, a linear energy scale is applied to convert the sum of the hit corrected energies (in PEs) into event energy (in keV). The conversion factor is estimated from the so-called double escape peak at 1592~keV from the 2615~keV \Tl{208} gamma line from a \Th{228} calibration source.

As shown in Fig.~\ref{fig:ebgmc}, the measured energy spectrum is featureless and monotonically decreasing for electron energies above 600~keV, with the rate decreasing by about one order of magnitude as the electron energy doubles (1200~keV). The distribution is the characteristic one of a beta spectrum, with no evidence for gamma-ray lines. Unlike in the calibration spectrum presented in \cite{Renner:2018ttw} where a fiducial volume cut is applied, gamma-induced electrons are suppressed in the figure by the requirement of a cathode origin for the track. The \Bi{214} MC energy spectrum shows similar characteristics. A variation in the ambient background between the E1 and E2 periods (data were taken without the lead castle isolating the detector) might be responsible for some distortions in the data spectrum. In particular, a variation of 9\% in the total background rate in the fiducial volume has been measured between both periods. Such a variation is assumed to be responsible for the energy structures observed in the 1400--1700 keV range. However, the resulting data-Monte Carlo differences are not expected to have a significant impact in the Rn-induced background extrapolations for the NEXT-100 detector.

\section{\label{sec:betabetabackgrounds}Implications for double beta decay searches}
Finally, we study the impact that radon-induced backgrounds are expected to have on the NEXT physics program. In the following, the implications for the two-neutrino double beta decay (\bbtwonu) measurement of \Xe{136} in NEXT-White and for the neutrinoless double beta decay (\bbnonu) search in NEXT-100 are discussed. NEXT-White and NEXT-100 detector dimensions are compared in Tab.~\ref{tab:nextwhite_vs_next100}. More details about NEXT-100 can be found in \cite{Alvarez:2012flf}.

\begin{table}[!htb]
  \caption{\label{tab:nextwhite_vs_next100}Comparison of NEXT-White and NEXT-100 detector geoemtries, and adopted/planned run configurations.}
\begin{center}
\begin{tabular}{ccc}
  \hline
  Detector & NEXT-White & NEXT-100 \\\hline
  Maximum drift length (m) & 0.530 & 1.300 \\
  Active region diameter (m) & 0.396 & 1.070 \\
  Active gas volume (m$^3$) & 0.065 & 1.169 \\
  Total (active + buffer) gas volume (m$^3$) & 0.081 & 1.286 \\
  Surface in contact with total gas volume (m$^2$) & 1.068 & 6.606 \\ \hline
  Gas pressure (bar) & 7.2--15 & 15 \\
  Xenon mass in active volume (kg) & 2.6--5.6 & 100.5 \\ \hline
\end{tabular}
\end{center}
\end{table}

For NEXT-White, \Bi{214} decays from the cathode have been simulated at 15~bar pressure, in anticipation of the operating pressure for the upcoming NEXT physics runs with \Xe{136}-enriched xenon. Only simulated events with a minimum deposited energy of 500~keV are kept for further processing. A fully realistic simulation has been performed, as described in Sec.~\ref{sec:radonelectrons}. Energy and hit reconstruction for simulated data are obtained by using the same algorithms that are applied to real data, as described in Sec.~\ref{sec:radonelectrons}. In addition, the following reconstruction and \bbtwonu selection steps are performed on the simulated \Bi{214} background events:

\begin{description}
\item[1 S1 and 1 S2:] Only events with one S1 peak, and one S2 peak, are kept.
\item[Non-zero active volume hits:] It is possible to have events where all hits are reconstructed for $Z>530.3$~mm, that is outside the active volume boundaries. This unphysical situation can occur in \Bi{214}-\Po{214} delayed coincidences, where the reconstructed S1 originates from the electron and the reconstructed S2 from the time-delayed alpha particle. We reject such events in order to suppress all alpha particle activity.
\item[Fiducial radius:] Events are required to satisfy $R_\mathrm{max}<178$~mm, where the maximum radial position is obtained from the entire collection of reconstructed hits in the event. In other words, no reconstructed hits within 20~mm from the active volume boundaries are allowed in the barrel region.
\item[Fiducial $Z$ position:] Events are required to satisfy $Z_\mathrm{min}>20$~mm, $Z_\mathrm{max}<512$~mm, where the minimum and maximum Z positions are obtained from the entire collection of reconstructed hits in the event. In other words, no reconstructed hits within 20~mm from the active volume boundaries are allowed in the two end-cap regions.
\item[Single track:] Once hits are reconstructed, they are first grouped into 3D volume elements (or voxels) of 10~mm size. Then, a “Breadth First Search” (BFS) algorithm \cite{Ferrario:2015kta} is applied to reconstruct tracks and to identify track extremes from a collection of connected voxels. Only events with a single reconstructed track are kept.
\item[Blob cut:] Energy ``blob'' candidates are built around the two track extremes. For each track extreme, if an energy greater than 300~keV within a radius of 15~mm is found, the extreme is considered to be a valid blob candidate. Only single-track events with two valid blob candidates are kept.
\item[Energy ROI:] The event energy $E_\mathrm{reco}$ is obtained by summing the hit charges after correcting for detector non-uniformities in response, and after applying an energy scale factor to convert PEs into keV. Only events within the energy region of interest (ROI), defined in the \bbtwonu case to be $E_\mathrm{reco}>700$~keV, are retained.
\end{description}

\begin{table}[!htb]
  \caption{\label{tab:bi214mc}Cumulative efficiencies through the various event selection criteria, for simulated cathode \Bi{214} decays in NEXT-White and NEXT-100 at 15~bar pressure. For NEXT-White (NEXT-100), the event selection corresponds to the \bbtwonu (\bbnonu) criteria.}
\begin{center}
\begin{tabular}{cccc}
\hline
Selection                   & \multicolumn{2}{c}{NEXT-White}              & NEXT-100 \\ \hline
                            & Full \bbtwonu        & Fast \bbtwonu        & Fast \bbnonu  \\
                            & analysis             & analysis             & analysis \\ \hline
None                        &       1              & 1                    & 1 \\
Min. deposited energy       & $1.54\times 10^{-1}$  & $1.54\times 10^{-1}$ & $1.59\times 10^{-2}$ \\
1 S1 and 1 S2               & $1.14\times 10^{-1}$  & N/A                  & N/A \\
Non-zero active volume hits & $1.13\times 10^{-1}$  & $1.50\times 10^{-1}$ & $1.42\times 10^{-2}$ \\
Fiducial radius             & $8.17\times 10^{-2}$  & $1.07\times 10^{-1}$ & $1.21\times 10^{-2}$ \\
Fiducial $Z$ position       & $5.76\times 10^{-3}$  & $2.81\times 10^{-3}$ & $8.03\times 10^{-4}$ \\
Single track                & $5.09\times 10^{-3}$  & $1.71\times 10^{-3}$ & $4.45\times 10^{-4}$ \\
Blob cut                    & $9.40\times 10^{-4}$  & $7.66\times 10^{-4}$ & $9.53\times 10^{-5}$ \\
Energy ROI                  & $8.62\times 10^{-4}$  & $3.00\times 10^{-4}$ & $4.00\times 10^{-8}$  \\ \hline
\end{tabular}
\end{center}
\end{table}

Table~\ref{tab:bi214mc} shows the \Bi{214} event reduction summary through the various reconstruction and selection steps described above. The above description corresponds to the NEXT-White ``Full \bbtwonu analysis'' column in Tab.~\ref{tab:bi214mc}. Table~~\ref{tab:bi214mc} also contains a NEXT-White ``Fast \bbtwonu analysis'' column computed via the same procedure as the one described in \cite{Martin-Albo:2015rhw}. In this case, the energy depositions within the gas are processed through a pseudo-reconstruction step to build event energies and reconstructed voxels, without starting from digitized waveforms as is the case in the full analysis. All selection cuts described above for the full analysis are also applicable to the fast analysis, with the exception of the ``1 S1 and 1 S2 `` requirement. After all cuts, the background acceptance from the full \bbtwonu analysis is about $8.6\times 10^{-4}$, and about 3 times lower from the more idealized fast analysis.

\begin{figure}[!htb]
  \begin{center}
    \includegraphics[width=0.49\textwidth]{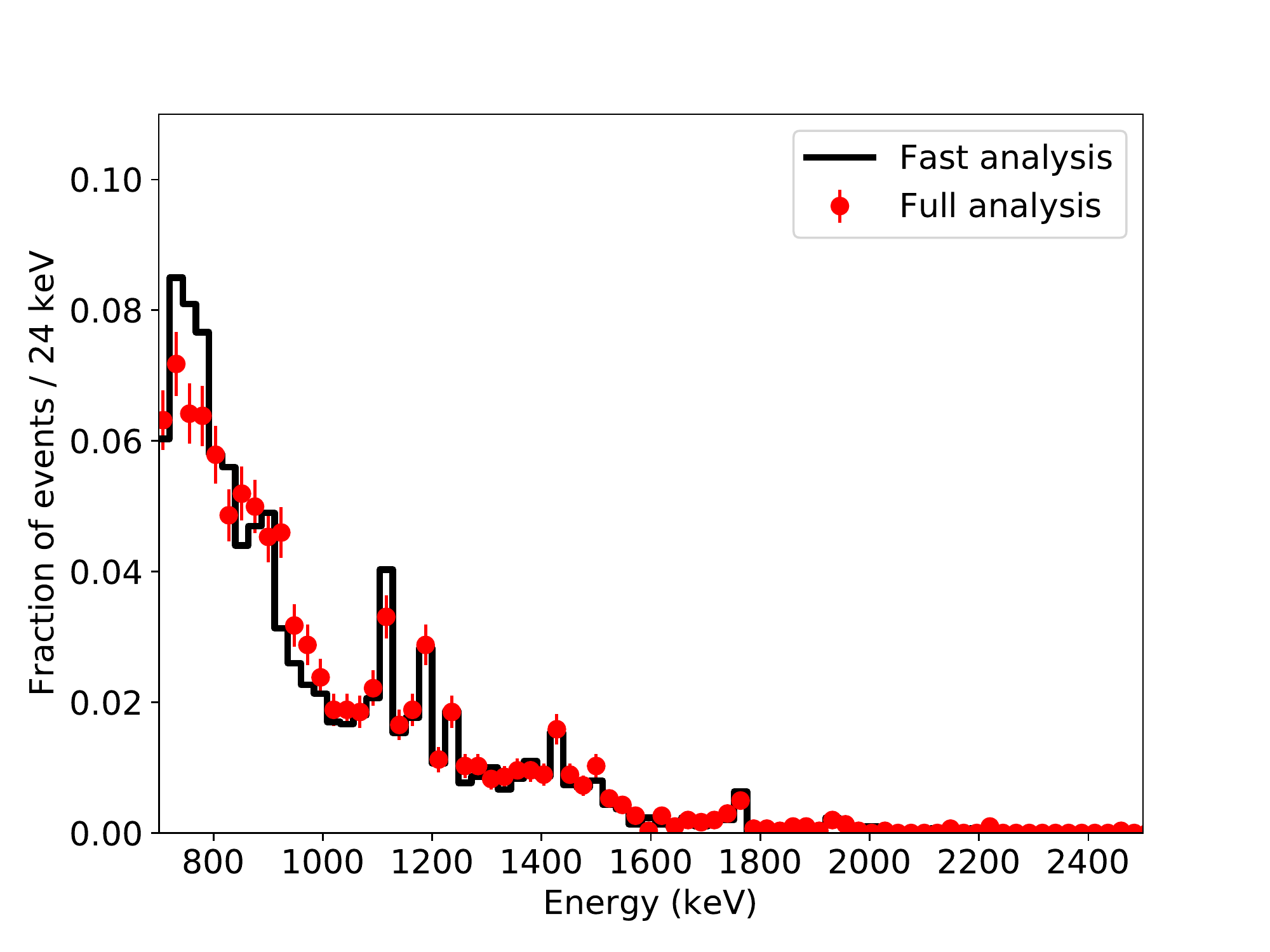} \hfill
    \includegraphics[width=0.49\textwidth]{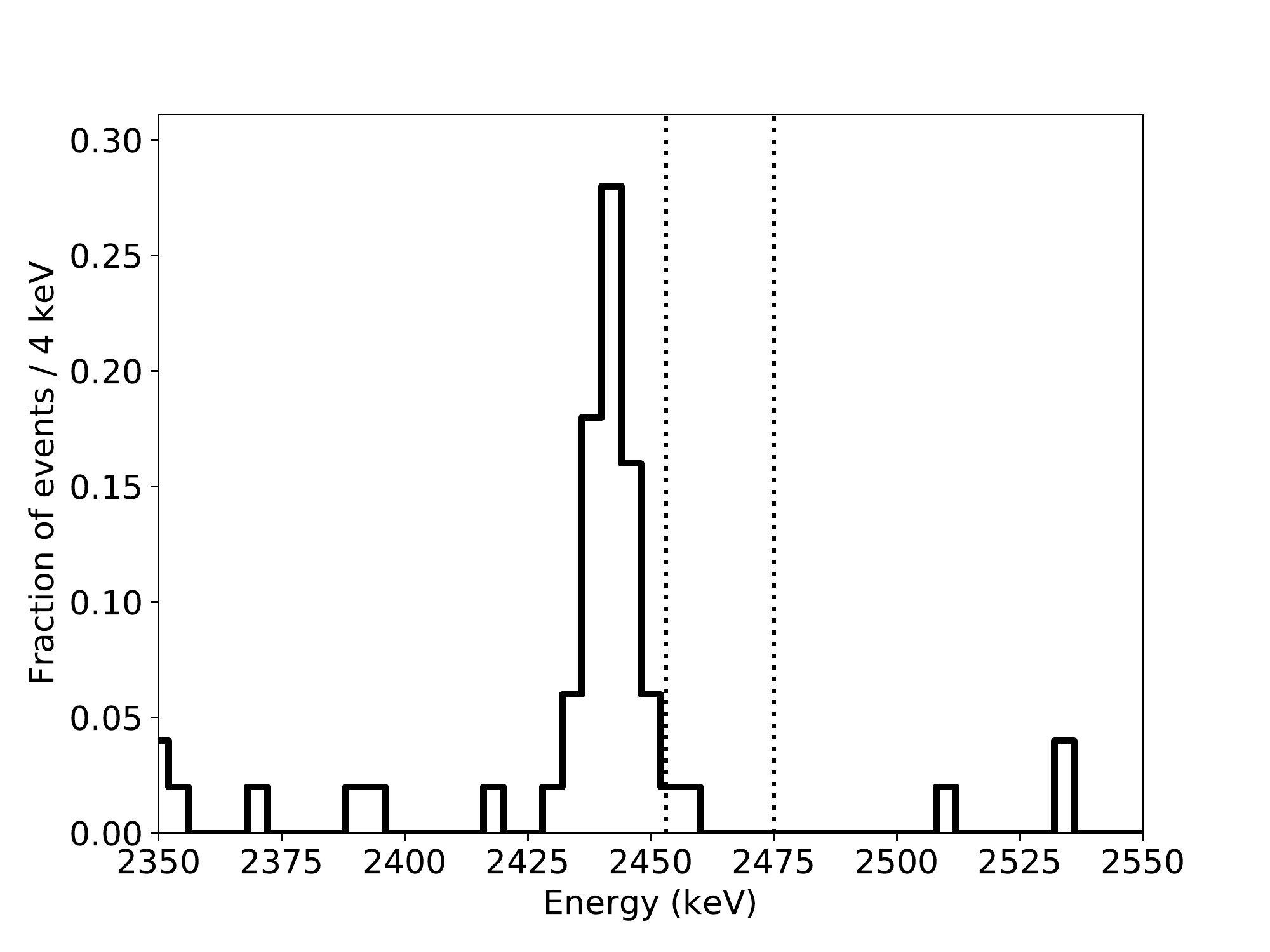} 
    \caption{\label{fig:bi214mc}Distributions of reconstructed event energy for cathode \Bi{214} simulated decays. Left panel: simulated events in NEXT-White at 15~bar and passing all \bbtwonu cuts. The black histograms show the expectations from the fast analysis, the red markers from the full analysis. Right panel: simulated events in NEXT-100 at 15~bar and passing all \bbnonu fast analysis cuts except the energy ROI one, indicated by the vertical dashed lines. All curves are unit-area normalized.}
  \end{center}
\end{figure}

The left panel of Fig.~\ref{fig:bi214mc} shows the reconstructed energy for NEXT-White \Bi{214} events simulated from the cathode, and passing all \bbtwonu cuts in Tab.~\ref{tab:bi214mc}. Unit-area normalized distributions for both the full and fast analyses are shown. The agreement between the fast and full analysis energy distribution shapes is excellent. A number of full energy deposition gamma-ray lines are clearly visible, at 768, 1120, 1238 and 1764~keV energies \cite{Wu:2009lpp}. In addition, the double-escape peaks at 1182 and 1426~keV from the 2204 and 2448~keV gamma-ray lines \cite{Wu:2009lpp}, respectively, are also visible.

From the \Rn{222} specific activity measurement during the low-radon alpha run, $(38.1\pm 6.3)$~mBq/m$^3$ as discussed in Sec.~\ref{sec:radonactivity}, the radon-induced background rate for the upcoming \bbtwonu measurement in NEXT-White has been estimated. We assume \Rn{222} concentration to be uniform throughout the full active (530.3~mm long) and buffer (130~mm long) volumes, with radius 198~mm. The total \Rn{222} activity is therefore $(3.11\pm 0.52)$~mBq. The same activity is assumed for \Bi{214} decays from the cathode. For a $8.6\times 10^{-4}$ background acceptance after cuts as obtained from the full \bbtwonu analysis in Tab.~\ref{tab:bi214mc}, a total radon-induced background rate of $(85\pm 14)$~counts/yr is obtained. This rate is about two orders of magnitude smaller than the background expected from \Co{60}, \K{40}, \Bi{214} and \Tl{208} decays altogether from radioactive impurities trapped in the NEXT-White detector components. Radon-induced backgrounds will therefore be negligible in NEXT-White.

Simulations have also been performed for cathode \Bi{214} decays in NEXT-100. The simulation, reconstruction and analysis steps follow the ones described above for NEXT-White, with two exceptions. First, only a fast \bbnonu analysis as in \cite{Martin-Albo:2015rhw} has been performed in this case, since the tools to perform a full analysis in NEXT-100 are still under development. Second, the event selection is optimized for the \bbnonu search. As such, a minimum deposited energy per event of 2300~keV is required for further processing, and the energy ROI is defined to be $2453<E_\mathrm{reco}<2475$~keV. All other reconstruction and selection steps are identical to the ones described above, including the fiducial requirement of no reconstructed hits within 20~mm of the detector active volume boundaries. Table~\ref{tab:bi214mc} gives also the event reduction summary of NEXT-100 \Bi{214} background events from the cathode and for the fast \bbnonu analysis. Primarily because of the stricter energy requirement, a much smaller background acceptance is obtained, $4\times 10^{-8}$. We can also use Tab.~\ref{tab:bi214mc} to approximately estimate the cathode \Bi{214} background rejection in NEXT-100 if a full \bbnonu analysis were available. To this end, we multiply the $4\times 10^{-8}$ background acceptance figure by the ratio of the NEXT-White full and fast analysis acceptances, to obtain an acceptance of about $1.1\times 10^{-7}$.

The right panel of Fig.~\ref{fig:bi214mc} shows the unit-area normalized energy distribution for \Bi{214} cathode events in NEXT-100 passing all \bbnonu cuts except the energy ROI one. The distribution is dominated by the 2448~keV gamma-ray line of \Bi{214}, located only 10~keV below the \Xe{136} $Q_{\beta\beta}$ value. Events in the high-energy tail of the 2448~keV reconstructed energy peak can pass the energy ROI selection, also shown in the figure with vertical lines.

\begin{table}[!htb]
\caption{\label{tab:next100backgroundrate}Radon-induced background rates expected in NEXT-100 under different assumptions. The \Rn{222}-induced total \Bi{214} activity from the cathode given in the table is the same as the one measured in NEXT-White during period A1 in the optimistic scenario, while it is assumed to scale as the detector surface area in the pessimistic scenario. The background acceptances and background rates refer to the \bbnonu selection.}
\begin{center}
\begin{tabular}{cccc}
\hline
Scenario & Total \Bi{214} activity from & Background  & \bbnonu background rate  \\ 
         & cathode (counts/yr)          & acceptance  & (counts/yr) \\ \hline
Optimistic  & $(9.8\pm 1.6)\times 10^4$ & $4\times 10^{-8}$ & $(3.9\pm 0.7)\times 10^{-3}$ \\
Pessimistic & $(6.1\pm 1.0)\times 10^5$ & $1.1\times 10^{-7}$ & $0.07\pm 0.01$ \\ \hline
\end{tabular}
\end{center}
\end{table}

Table~\ref{tab:next100backgroundrate} shows the radon-induced background rate expected in NEXT-100 \bbnonu searches, again starting from the NEXT-White radon activity measurement in Sec.~\ref{sec:radonactivity}, and the background acceptance numbers quoted above. The background prediction is more uncertain in this case, for two reasons. First, as we have not identified the radon source in the NEXT-White measurement, some uncertainty exists on how the total \Rn{222} activity extrapolates from NEXT-White to NEXT-100. In the optimistic scenario, radon emanation is dominated by gas system components external to the detector vessels. As the same gas system is foreseen for NEXT-White and NEXT-100 operations, in this case the total \Bi{214} activity from the TPC cathodes would be the same in both detectors. This total activity has been measured to be $(9.8\pm 1.6)\times 10^4$ counts/yr in NEXT-White during the A1 period. In the pessimistic scenario, radon emanation is dominated by inner detector components, and the radon total content would scale with detector surface area, about 6 times larger in NEXT-100 compared to NEXT-White, see Tab.~\ref{tab:nextwhite_vs_next100}. Therefore, a total \Bi{214} activity from the NEXT-100 cathode of $(6.1\pm 1.0)\times 10^5$ counts/yr is assumed in the pessimistic scenario. Second, the lack of full analysis results in NEXT-100 introduces an additional uncertainty arising from the background rejection performance for this event topology. For the optimistic scenario, we take the $4\times 10^{-8}$ background acceptance number given above. For the pessimistic scenario, we assume $1.1\times 10^{-7}$. Despite the fact the former number is not based on a full simulation and reconstruction, we consider it to be a realistic estimate of the best possible background rejection performance, as more detailed simulations in the future will be accompanied by improvements in event reconstruction and selection.

As a result, we define two extreme scenarios in Tab.~\ref{tab:next100backgroundrate}. The optimistic one assumes the lowest possible radon content in the detector, and the best possible background rejection. A radon-induced background rate of only $(3.9\pm 0.7)\times 10^{-3}$~counts/yr is expected in this case. The pessimistic scenario assumes high radon content, and the worst background rejection number. The background rate is expected to be $0.07\pm 0.01$~counts/yr in the pessimistic scenario. These numbers should be compared with the background rate expected from radioactive impurities in detector materials, estimated to be at the level of $4\times 10^{-4}$~counts/(keV$\cdot$kg$\cdot$yr), or 1.06~counts/yr \cite{Martin-Albo:2015rhw}. Hence, the NEXT-White measurement allows us to constrain the radon-induced background to be at most one order of magnitude smaller than the detector radioactive impurities, and possibly much smaller. In any case, we can conclude that radon-induced backgrounds will remain at a tolerable level for the NEXT-100 physics program. A campaign to quantify radon emanation from NEXT detector and gas system components is underway, to identify the dominant radon sources and hence to further reduce the uncertainties in the extrapolation from the NEXT-White measurement to NEXT-100.

\section{\label{sec:conclusions}Conclusions}

The background activity induced by radon, particularly \Rn{222}, has proven to be a serious concern for most double beta decay experiments. This work describes the measurements of the internal \Rn{222} activity in NEXT-White during the so-called Run-II period, using data collected between March and August 2017. The detector was filled with \Xe{136}-depleted xenon gas during this period, permitting a first validation in NEXT of one of the major background sources in double beta decay searches. This is the first data-driven study by the NEXT Collaboration to measure and understand absolute event rates, and hence to understand efficiencies for event triggering, reconstruction and selection. Both radon-induced alpha particles and radon-induced electrons have been studied. Implications of these observations for two-neutrino and neutrinoless double beta decay searches of \Xe{136}, in NEXT-White and NEXT-100, respectively, are discussed.

The radon activity is measured through the alpha production rate induced in the fiducial volume by \Rn{222} and its alpha-emitting progeny. Three alpha run periods are considered. A first (A1) period corresponds to a time when the ambient temperature getter, known to be an intense radon emitter, has not yet been turned on after closing the detector. This is the period used to estimate the \Rn{222} internal activity for the upcoming physics runs in NEXT-White and NEXT-100 with \Xe{136}-enriched xenon. A second (A2) period is used to quantify how radon activity decreases over time after ambient temperature getter operations in NEXT-White. The alpha production rate measured during A2 is found to decrease with a half-life of $T_{1/2}=(3.871\pm 0.013)~d$, once DAQ dead-time corrections are accounted for. A third (A3) period allows us to estimate the rate of radon-induced electrons emitted from the cathode from the decay of \Bi{214}, progeny of \Rn{222}, for electron data taken in the same period as the A3 alpha data.

The energy released in alpha decays is reconstructed, in order to identify the alpha-emitting isotopes. The energy spectrum during the high \Rn{222} activity period A3 shows three alpha populations within the detector fiducial volume, all from the same \Rn{222} decay chain: \Rn{222} (5590~keV) itself, \Po{218} (6115~keV) and trace amounts of \Po{214} (7834~keV). In the low radon A1 period, only the \Rn{222} and \Po{218} isotopes are clearly identified. We conclude that essentially all of the \Po{214} decays induced by internal \Rn{222} originate from the cathode plane. Owing to the short \Po{214} half-life, the same assumption is made for the spatial distribution of the electron-emitting \Bi{214} parent ions. A \Rn{222} specific activity of $(38.1\pm 2.2\mathrm{(stat.)}\pm 5.9\mathrm{(syst.)})$~mBq/m$^3$ is obtained during the low \Rn{222} activity period A1. The \Rn{222} activity in the same fiducial volume increases by three orders of magnitude during A3, taken only six days after closing the \Rn{222}-emanating ambient temperature getter.

Radon-induced electrons have been studied during NEXT-White Run-II, through background runs taken with no calibration sources and nominal electric fields. A first (E1 or high \Rn{222} activity) electron period corresponds to 5 days after turning off the ambient temperature getter. A second (E2 or low \Rn{222} activity) period corresponds to data taken 16 days after E1. The radon-induced population of electrons is clearly visible in the average drift length ($Z$) distribution of electron events, with a large event excess at the cathode position. The rate of cathode electrons decreases with time as radon decays, confirming their interpretation as cathode \Bi{214} decay events induced by radon. The detailed topologies observed for $>1.5$~MeV electron events produced at large drift during the high \Rn{222} activity period are consistent with this interpretation. The energy spectrum of the electron background above 600~keV has been reconstructed. The energy distribution shows a monotonically decreasing and featureless beta spectrum, with no evidence for gamma-ray lines. The Run-II electron distributions are compared to a Monte-Carlo (MC) simulation of $^{214}$Bi decays from the cathode. The MC normalization is estimated from the high \Rn{222} alpha dataset A3, taken during the same period as the electron data. We conclude that it is possible to infer the rate of \Bi{214} electrons emitted from the cathode by monitoring the alpha production rate within the fiducial volume, within the quoted systematic uncertainty of about 20\%.

Expectations of radon-induced double beta decay backgrounds for NEXT have been updated according to this new measurement. The impact of radon-induced backgrounds has been estimated in both NEXT-White and NEXT-100 with simulated data and fast analysis tools identical to the ones used in \cite{Martin-Albo:2015rhw}. In addition, a full \bbtwonu analysis in NEXT-White has been performed, relying on a complete simulation of detector response including drift, optical and electronics effects, and by reconstructing simulated events in the same way as real data. For the NEXT-White full \bbtwonu analysis, the \Rn{222}-induced background rate is expected to be $(85\pm 14)$~counts/yr for events passing all \bbtwonu cuts. This rate is about two orders of magnitude smaller than the background expected from \Co{60}, \K{40}, \Bi{214} and \Tl{208} decays altogether from radioactive impurities trapped in the detector components. The radon-induced contribution to the background rate for \bbnonu searches in NEXT-100 has been quantified. The rate has been estimated under different assumptions for the extrapolation of the \Rn{222} activity measurement in NEXT-White to NEXT-100, and for the achievable rejection performance of cathode \Bi{214} background events. Even in the pessimistic scenario, with highest \Rn{222} activity and poorest background rejection, the radon-induced background is expected to be $0.07\pm 0.01$~counts/yr. This rate is at most one order of magnitude smaller than the rate induced by radioactive impurities trapped in detector materials, and possibly much smaller. Since radon will become a more serious issue for a next-generation ton-scale detector based on the NEXT technology, the installation of an active filtration system to mitigate internal radon (a radon trap) is being considered by the NEXT Collaboration.

\acknowledgments
The NEXT Collaboration acknowledges support from the following agencies and institutions: the European Research Council (ERC) under the Advanced Grant 339787-NEXT; the European Union’s Framework Programme for Research and Innovation Horizon 2020 (2014-2020) under the Marie Skłodowska-Curie Grant Agreements No. 674896, 690575 and 740055; the Ministerio de Econom\'ia y Competitividad of Spain under grants FIS2014-53371-C04, the Severo Ochoa Program SEV-2014-0398 and the Mar\'ia de Maetzu Program MDM-2016-0692; the GVA of Spain under grants PROMETEO/2016/120 and SEJI/2017/011; the Portuguese FCT and FEDER through the program COMPETE, projects PTDC/FIS-NUC/2525/2014 and UID/FIS/04559/2013; the U.S.\ Department of Energy under contracts number DE-AC02-07CH11359 (Fermi National Accelerator Laboratory), DE-FG02-13ER42020 (Texas A\&M) and DE-SC0017721 (University of Texas at Arlington); the University of Texas at Arlington; and the Foundation for Polish Science (Grant No. TEAM/2016-2/17). We also warmly acknowledge the Laboratorio Nazionale di Gran Sasso (LNGS) and the Dark Side collaboration for their help with TPB coating of various parts of the NEXT-White TPC. Finally, we are grateful to the Laboratorio Subterr\'aneo de Canfranc for hosting and supporting the NEXT experiment.

\bibliographystyle{JHEP}
\bibliography{biblio}

\end{document}